\documentclass[12pt]{article}

\usepackage{amssymb}
\usepackage{amsmath}
\usepackage{array} 
\usepackage{epsfig}
\usepackage{graphics}

\begin{document}

\begin{center}

 {\Large \bf
\vskip 7cm
\mbox{Elastic $\pi^+ p$  and $\pi^+\pi^+$ scattering at LHC.}
}
\vskip 1cm

\mbox{A.E.~Sobol*, R.A.~Ryutin*, V.A.~Petrov*, }\\
\vskip 0.5cm
\mbox{*{\small Institute for High Energy Physics}}
\mbox{{\small{\it 142 281} Protvino, Russia}}\\
\vskip 0.5cm
\mbox{M.~Murray**}\\
\vskip 0.5cm
\mbox{**{\small University of Kansas, USA}}

 \vskip 1.75cm
{\bf
\mbox{Abstract}}
  \vskip 0.3cm

\newlength{\qqq}
\settowidth{\qqq}{In the framework of the operator product  expansion, the quark mass dependence of}
\hfill
\noindent
\begin{minipage}{\qqq}
We discuss the possibility of  measuring leading neutron production at the LHC.  These data could be used to extract from it $\pi^+ p$ and $\pi^+\pi^+$ cross-sections. In this note we give some estimates for the case of elastic cross-sections and discuss related problems and prospects.

\end{minipage}
\end{center}


\begin{center}
\vskip 0.5cm
{\bf
\mbox{Keywords}}
\vskip 0.3cm

\settowidth{\qqq}{In the framework of the operator product  expansion, the quark mass dependence of}
\hfill
\noindent
\begin{minipage}{\qqq}
Leading Neutron Spectra -- Elastic cross-section -- Absorption -- Regge-eikonal model
\end{minipage}

\end{center}

\setcounter{page}{1}
\newpage

\section{Introduction}
\begin{center}
 \begin{figure}[ht!]
 \hskip 1cm \vbox to 4cm
 {\hbox to 13cm{\epsfxsize=13cm\epsfysize=4cm\epsffile{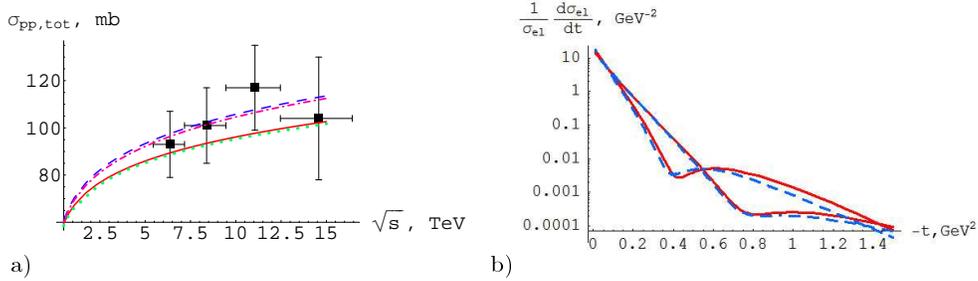}}}
 \caption{\small\it\label{fig:M1} a) Total $pp$ cross-sections in the energy range $0.5\;{\rm TeV}<\sqrt{s}<15\;{\rm TeV}$ for parametrizations~\cite{landshofftot} (solid),\cite{COMPETE} (dashed),\cite{BSW} (dotted) and~\cite{godizov1},\cite{godizov2} (dash-dotted). Data points are taken from~\cite{datapptot}; b) evolution of the diffractive pattern in the $pp$ scattering for $\sqrt{s}=1\to 20\;{\rm TeV}$ (from right to left) for parametrizations from~\cite{BSW} (solid) and~\cite{3Pomerons} (dashed).}
 \end{figure}
\end{center}

In a recent paper~\cite{ourneutrontot} we pushed forward 
the idea of using the Zero
Degree Calorimeters,  ZDCs, designed for different uses at 
several of the LHC
collaborations, to extract the total cross-section of the processes  $\pi p\rightarrow X$ and $\pi\pi\rightarrow X$ at energies about 1-5 TeV.  This would allow us to use the LHC as a $\pi p$ and $\pi\pi$ collider. In Ref.~\cite{ourneutrontot} there was also mentioned possible measurements of the elastic $\pi p$ and $\pi\pi$ scattering. The physics motivation for extending this program to elastic scattering is very clear since the total and elastic cross-sections are so tightly interrelated (e.g., via unitarity) that any testing of various models of high-energy hadron interactions is little informative use without both of them (Fig.~\ref{fig:M1}).

\begin{center}
 \begin{figure}[b!]
  \vbox to 4.5cm
 {\hbox to 15.5cm{\epsfxsize=15.5cm\epsfysize=4.5cm\epsffile{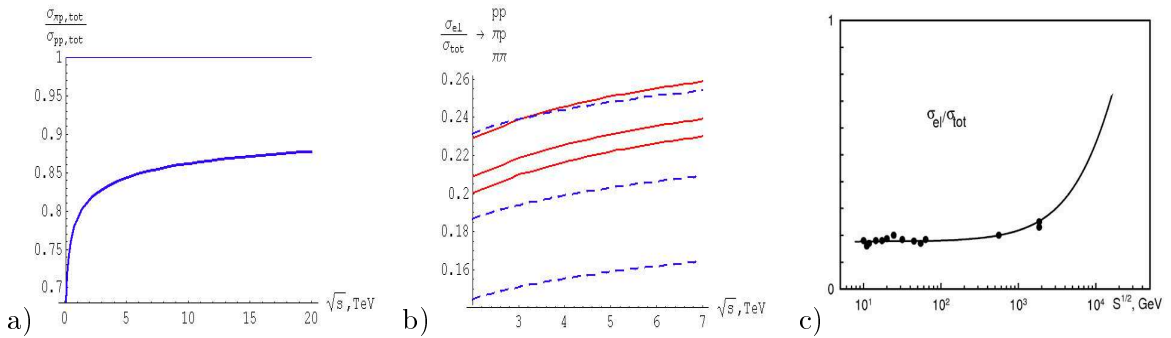}}}
 \caption{\small\it\label{fig:M2} a) Ratio of $\pi^+ p$ to $pp$ total cross-sections for COMPETE~\cite{COMPETE} parametrization; b) ratio of elastic to total cross-sections for $pp$, $\pi^+p$ and $\pi^+\pi^+$ processes for parametrizations~\cite{BSW} (solid) and~\cite{godizov1},\cite{godizov2} (dashed); c) ratio of proton-proton elastic to total cross-sections in~\cite{troshin1},\cite{troshin2}.}
 \end{figure}
\end{center}
\vspace*{-2cm}
 We would like also to remind to the reader an idea of \emph{universality of 
 strong interactions} at superhigh energies,in the sense that any ratio of two 
 total or elastic cross-sections will asymptotically approach 1
 independently of initial states (Fig.~\ref{fig:M2}a). How close we are to reaching the asymptotic regime, \emph{asymptopia},  can be tested when looking at
 the ratio of elastic to total cross-sections. Most of theories predict it to be 1/2~\cite{landshofftot}-\cite{godizov2} (Fig.~\ref{fig:M2}b) though there are some  
  theories which predict it to be equal to 1~\cite{troshin1},\cite{troshin2} (Fig.~\ref{fig:M2}c). Sure, this can and will be done. The TOTEM experiment has been designed to measure this ratio but any information about  such a feature for the interaction of the \emph{lightest
 hadrons} is impossible to overestimate. 

\begin{center}
 \begin{figure}[t!]
  \vbox to 4.5cm
 {\hbox to 14cm{\epsfxsize=14cm\epsfysize=4.5cm\epsffile{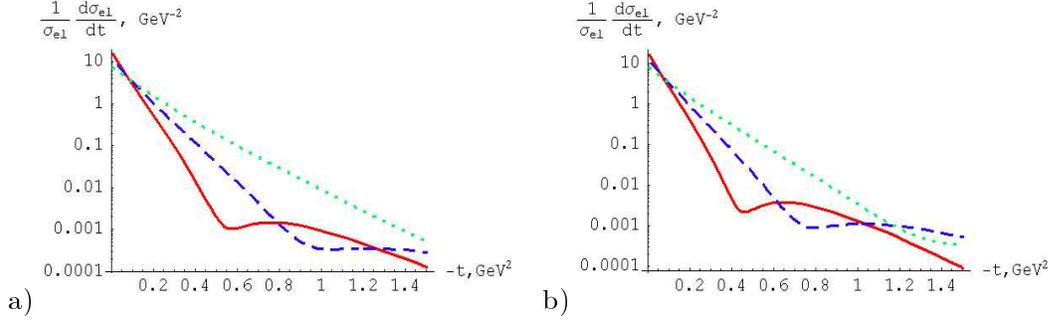}}}
 \caption{\small\it\label{fig:M3} Evolution of diffractive pattern for $pp$ (solid), $\pi^+p$ (dashed) and $\pi^+\pi^+$ (dotted) elastic processes at a) $\sqrt{s}=5$~TeV and b) $\sqrt{s}=14$~TeV for the parametrization from the Appendix B.}
 \end{figure}
\end{center} 
 \vskip -1.3cm
 There are many other
 questions to be asked. For instance, how different are the interaction
 radii in $pp$,$\pi p$ and $\pi\pi$ high-energy collisions? The properties of the interaction region could  be obtained from diffractive patterns, which are different for these
 processes at the same energy (Fig.~\ref{fig:M3}). It would be interesting to know the dependence of these interactions on the very different quark-gluon content of colliding particles etc. If it will be
 possible to provide such a marvellous opportunity as access to the
 data on $\pi p$ and $\pi\pi$ TeV-energy interactions all the landscape of
 the ``soft physics'' will be transformed to the better.
 In what follows we present our estimates on the possibility to
 extract $\pi p$ and $\pi\pi$ elastic cross-sections using CMS as an example. Just before sending this article to arxiv a paper on the four-body
 reaction $pp\to nn\pi^+\pi^+$ at the LHC was placed there. However, the authors of~\cite{poland2010} do not estimate possibilities for extraction of the $\pi\pi$ elastic cross-section.

\section{Exclusive single pion exchange}

\begin{center}
\begin{figure}[t!]
\hskip  4cm \vbox to 7cm
{\hbox to 7cm{\epsfxsize=7cm\epsfysize=7cm\epsffile{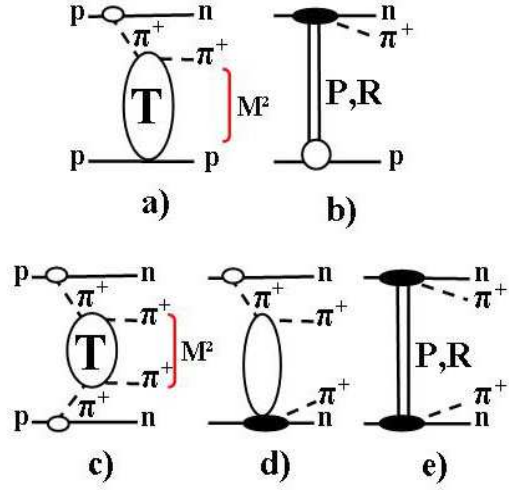}}}
\caption{\small\it\label{fig:1} Diagrams for the exclusive signal and background processes in leading neutrons production (initial rescattering corrections are not shown). a)~Signal for the elastic $\pi^+ p$ scattering: process with an exclusive single pion exchange (S$\pi$E) $p+p\to n+\pi^++p$, $M$ is the mass of the $\pi^+ p$ system; b)~background for the elastic $\pi^+ p$ scattering: low mass single dissociation with Pomeron and reggeon exchanges; c)~signal for the elastic $\pi^+\pi^+$ scattering: process with an exclusive double pion exchange (D$\pi$E) $p+p\to n+\pi^++\pi^++n$, $M$ is the mass of the $\pi^+ \pi^+$ system; d)~background for the elastic $\pi^+ \pi^+$ scattering: S$\pi$E with single low mass dissociation in the $\pi^+ p$ channel; e)~background for the elastic $\pi^+ \pi^+$ scattering: low mass double dissociation with Pomeron and reggeon exchanges.}
\end{figure}
\end{center}

 The diagram of the exclusive single-pion exchange (S$\pi$E) process $p+p\to n+\pi^++p$ is presented in Fig.~\ref{fig:1}a. The momenta are $p_1$, $p_2$, $p_n$, $p_{\pi}^{\prime}$, $p_2^{\prime}$ respectively. In the center-of-mass frame these can be represented as follows  (arrows denote transverse momenta):
 \begin{eqnarray}
&& \label{kin1a}p_i\simeq\left(\frac{\sqrt{s}}{2},(-1)^{i-1}\frac{\sqrt{s}}{2} ,\vec{0}\right),
p_n\simeq\left((1-\xi)\frac{\sqrt{s}}{2},(1-\xi)\frac{\sqrt{s}}{2},-\vec{q}\right),\\
&& \label{kin1b}p_2^{\prime}\simeq\left( (1-\xi_p)\frac{\sqrt{s}}{2},-(1-\xi_p)\frac{\sqrt{s}}{2},\vec{q}-\vec{k} \right),\\
&& \label{kin1c}p_{\pi}=p_1-p_n,\; p_{\pi}^{\prime}=p_2+p_{\pi}-p_2^{\prime},\\
&& \label{kin1e}M^2=(p_{\pi}+p_2)^2=(p_{\pi}^{\prime}+p_2^{\prime})^2,\\
&& \label{kin1f} t=(p_1-p_n)^2=p_{\pi}^2\simeq-\frac{\vec{q}^{\; 2}+\xi^2m_p^2}{1-\xi},\; \xi\simeq\frac{M^2}{s},\\
&& t_p=(p_2-p_2^{\prime})^2\simeq-\frac{(\vec{q}-\vec{k})^2+\xi_p^2m_p^2}{1-\xi_p},\; \xi_p\simeq\frac{m_{\pi}^2+\vec{k}^{\;2}}{\xi\; s},
 \end{eqnarray}
\begin{center}
\begin{figure}[t!]
\hskip  3cm \vbox to 5cm
{\hbox to 9cm{\epsfxsize=9cm\epsfysize=5cm\epsffile{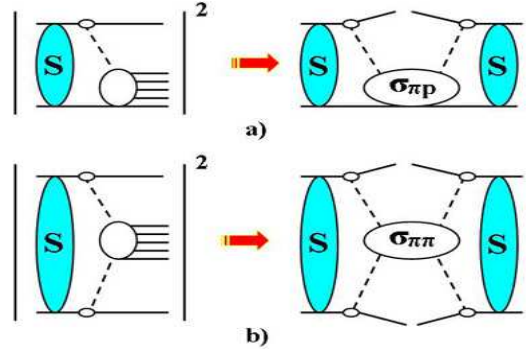}}}
\caption{\small\it\label{fig:2} Amplitudes squared and cross-sections of the processes: a) $p+p\to n+X$ (S$\pi$E), b) $p+p\to n+X+n$ (D$\pi$E). $S$ represents soft rescattering corrections. In this note $X=\pi^++p$ (elastic S$\pi$E) and $X=\pi^+ +\pi^+$ (elastic D$\pi$E).}
\end{figure}
\end{center}
\begin{center}
\vskip -2.5cm
\end{center}
As an approximation for $\pi$ exchange we use the formula shown graphically in Fig.~\ref{fig:2}a). If we take into account absorptive corrections this formula can be rewritten as 
\begin{equation}
\label{born}
\frac{d\sigma_{X,{\rm S}\pi {\rm E}}}{d\xi dt\; d\Phi_X}=\frac{G_{\pi^+pn}^2}{16\pi^2}\frac{-t}{(t-m_{\pi}^2)^2} F^2(t) \xi^{1-2\alpha_{\pi}(t)} \frac{d\sigma_{X,\pi^+ p}(\xi s)}{d\Phi_X} S(s/s_0, \xi, t),
\end{equation}
where $\Phi_X$ is the phase space for the system X produced in the $\pi^+ p$ scattering, the pion trajectory is $\alpha_{\pi}(t)=\alpha^{\prime}_{\pi}(t-m_{\pi}^2)$. The slope $\alpha^{\prime}\simeq 0.9$~GeV$^{-2}$, $\xi=1-x_L$, were $x_L$ is the fraction of the initial proton's longitudinal momentum carried by the neutron, and $G_{\pi^0pp}^2/(4\pi)=G_{\pi^+pn}^2/(8\pi)=13.75$~\cite{constG}. The form factor $F(t)$ is usually expressed as an exponential
\begin{equation}
\label{formfactor}
F(t)=\exp(bt),
\end{equation}
where, from recent data~\cite{HERA2},\cite{KMRn1c16}, we expect $b\simeq 0.3\; {\rm GeV}^{-2}$.
 We are interested in the kinematical range $0.01$~GeV$^2<|t|<0.5$~GeV$^2$, $\xi<0.4$, where formula~(\ref{born}) dominates according to~\cite{KMRn1c13} and~\cite{KMRn1c14}. At high energies we can use any adequate parametrizations of different $\pi^+ p$ cross-sections. Here we replace $d\sigma_{X,\pi^+ p}/d\Phi_X$ in~(\ref{born}) by $d\sigma_{{\rm el},\pi^+p}/dt_p$ or integrated $\sigma_{{\rm el},\pi^+p}$ instead of $\sigma_{{\rm tot},\pi^+p}$ in~\cite{ourneutrontot}.

 The suppression factor $S$ arises from absorptive corrections~\cite{KMRn1}. We estimate absorption in the initial state for inclusive reactions and for both initial and final states in exclusive exchanges. For this task we use our model with 3 Pomeron trajectories~\cite{3Pomerons}:
\begin{eqnarray}\label{3IPtrajectories}
\alpha_{IP_1}(t)-1&=& (0.0578\pm0.002)+(0.5596\pm0.0078)t \;,\nonumber\\
\alpha_{IP_2}(t)-1&=& (0.1669\pm0.0012)+(0.2733\pm 0.0056)t \;,\nonumber\\
\alpha_{IP_3}(t)-1&=& (0.2032\pm0.0041)+(0.0937\pm0.0029)t \;,
\end{eqnarray}
These trajectories  are the result of a 20 parameter fit of the total and differential cross-sections in the region 
$$
0.01\;{\rm GeV}^2<|t|<14\; {\rm GeV}^2,\; 8\; {\rm GeV}<\sqrt{s}<1800\; {\rm GeV}.
$$
Although the $\chi^2/d.o.f.=2.74$ is rather large, the model gives
good predictions for the elastic scattering (especially in the low-t region with $\chi^2/d.o.f.\sim1$). 

We use the procedure described in~\cite{workn1},\cite{workn2} to estimate the absorptive corrections. With an effective factorized form of the, here-under, expression~(\ref{Udsigma}) used for convenience, we obtain:
\begin{equation}
 \label{Udsigma}\frac{d\sigma_{{\rm el},\; {\rm S}\pi{\rm E}}(s/s_0,\xi,\vec{q}^{\;2})}{d\xi d\vec{q}^{\;2}dt_p}=(m_p^2\xi^2+\vec{q}^{\;2})|\Phi_B(\xi,\vec{q}^{\;2})|^2\frac{\xi}{(1-\xi)^2}\frac{d\sigma_{{\rm el},\;\pi^+p}(\xi\;s)}{dt_p}S(s/s_0,\xi,\vec{q}^{\;2}),
\end{equation}
\begin{equation}
 \label{survival}S=\frac{m_p^2\xi^2 |\Phi_0(s/s_0,\xi,\vec{q}^{\;2})|^2+\vec{q}^{\;2}|\Phi_s(s/s_0,\xi,\vec{q}^{\;2})|^2}{(m_p^2\xi^2+\vec{q}^{\;2})|\Phi_B(\xi,\vec{q}^{\;2})|^2}.
\end{equation}
The functions $\Phi_0$ and $\Phi_s$ arise from different spin contributions to the amplitude
\begin{equation}
\label{spinamplitude}
A_{p\to n}=\frac{1}{\sqrt{1-\xi}}\bar{\Psi}_n \left(
m_p\xi\; \hat{\sigma}_3\cdot \Phi_0+\vec{q}\;\hat{\vec{\sigma}}\cdot\Phi_s
\right) \Psi_p
\end{equation}
and both are equal to $\Phi_B$ in the Born approximation. Here $\hat{\sigma}_i$ are Pauli matrices and $\bar{\Psi}_n$, $\Psi_p$ are neutron and proton spinors. Functions  $\Phi_{0,s,B}$ are given in the Appendix A. For the $\pi^+ p$ elastic cross-sections we use parametrizations from~\cite{BSW} and~\cite{godizov1},\cite{godizov2} which are described in Appendices B and C correspondingly.

\begin{center}
\begin{figure}[t!]
\hskip  0.5cm \vbox to 4.5cm
{\hbox to 12cm{\epsfxsize=12cm\epsfysize=4.5cm\epsffile{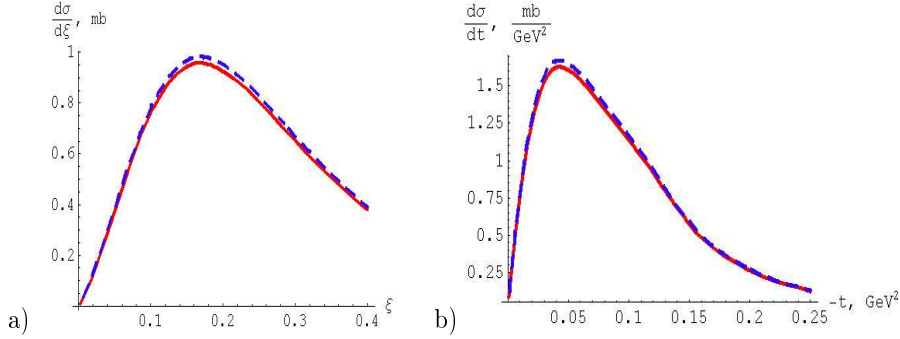}}}
\vskip 0.2cm
\caption{\small\it\label{fig:SpiEcs10TeV} Integrated cross-sections of the S$\pi$E process $p+p\to n+\pi^++p$ for parametrizations from Appendices B (solid) and C (dashed). a) $\frac{d\sigma}{d\xi}$ ($0.01\;{\rm GeV}^2<|t|<0.5\;{\rm GeV}^2$); b) $\frac{d\sigma}{dt},\; 10^{-3}<\xi<0.3$.}
\end{figure}
\end{center}

 The differential cross-sections for the process $p+p\to n+\pi^++p$ at $\sqrt{s}=10$~TeV are depicted in Fig.~\ref{fig:SpiEcs10TeV}. The total cross sections are listed in Table~\ref{tab:tot10TeVnum}. They are in the range  8-270 $\mu$b  for all values of $\xi_{max}<0.4$ implying that we will have plenty of rate for the measurements. 
\begin{table}[h!]
\caption{\small\it\label{tab:tot10TeVnum}Total $p+p\to n+\pi^++p$ cross-sections in the kinematical region $0<|\vec{q}|<0.5\;{\rm GeV}$,  $\xi_{min}=10^{-3}<\xi<\xi_{max}$ for two parametrizations given in Appendices B (C).}
\begin{center}
\begin{tabular}{|c|c|c|c|c|c|}
\hline
 $\xi_{max}$  &  0.05   & 0.1   & 0.2 & 0.3 & 0.4\\
\hline
 $\sigma_{p+p\to n+\pi^++p}$ , $\mu$b  &    8.5 (8.8)   &   37.5 (39)  &   128 (132)  & 208 (214) & 259 (266)\\
\hline
\end{tabular}
\end{center}
\end{table}

\begin{center}
\vskip -1.6cm
\end{center}
 At low energies ($\sqrt{s}<70$~GeV) the region of applicability of our  model is  given by the inequalities
 \begin{equation}
 \label{regdif}
 0.01\; {\rm GeV}^2<|t|<0.5\; {\rm GeV}^2,\; 10^{-6}<\xi<0.4. 
 \end{equation}
 At higher energies this region may be smaller (say $\xi<0.2$), since this corresponds to masses $M=3$~TeV at $\sqrt{s}=10$~TeV, and for larger masses the approach may break down.

\section{Exclusive double pion exchange}

\begin{center}
\begin{figure}[t!]
\hskip  1.5cm \vbox to 6.5cm
{\hbox to 10cm{\epsfxsize=10cm\epsfysize=6.5cm\epsffile{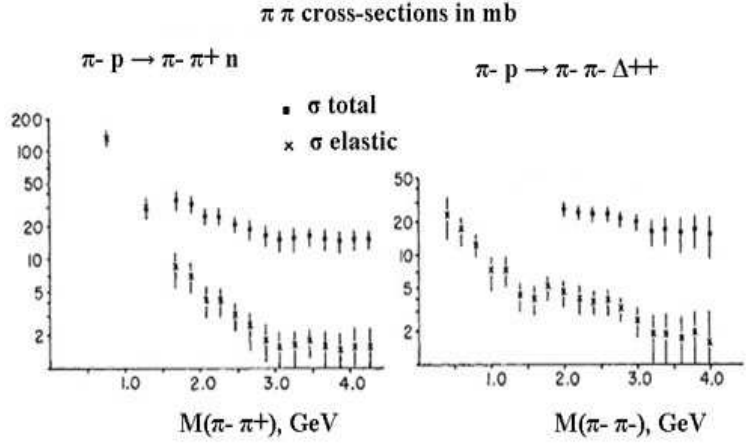}}}
\vskip 0.1cm
\caption{\small\it\label{fig:pipics} Elastic and total cross-sections for $\pi^-\pi^+$ and $\pi^-\pi^-$ scattering from the data on exclusive reactions as a function of the dipion invariant mass (Fig.5 from Ref.~\cite{pipiextract2}).}
\end{figure}
\end{center}

 As noted above, the Double pion Exchange (D$\pi$E) process can give information on both total and elastic $\pi\pi$ sross-sections. $\pi\pi$ cross-sections have been extracted in the past using the exclusive cross-section~\cite{pipiextract2}. The results  are shown in Fig.~\ref{fig:pipics}. There is some tendency for an early flattening of the $\pi\pi$ cross-sections. In $\pi p$ and $pp$ cross-sections this flattening begins at higher energies and precedes further growth. 
\begin{center}
\vskip -1.9cm
\end{center} 

We can extend the analysis for one pion exchange described above to double pion exchange (Fig.~\ref{fig:2}b, D$\pi$E).  The kinematics of the exclusive D$\pi$E ($p+p\to n+\pi^++\pi^++n$, the momenta are  $p_1$, $p_2$, $p_{n_1}$, $p_{\pi_1}^{\prime}$, $p_{\pi_2}^{\prime}$, $p_{n_2}$ respectively) is similar to the exclusive double pomeron exchange process: 
  \begin{equation}
 \label{kindpie1}p_{\pi_i}\simeq\left( \xi_i\frac{\sqrt{s}}{2},(-1)^{i-1}\xi_i\frac{\sqrt{s}}{2},\vec{q}_i\right),\; p_{n_i}=p_i-p_{\pi_i}
\end{equation}
\begin{eqnarray}
 && \label{kindpie5}M^2=(p_{\pi_1}+p_{\pi_2})^2\simeq\xi_1\xi_2 s -\left(\vec{q}_1+\vec{q}_2\right)^2\simeq \xi_1\xi_2 s,\\
&&  \label{kindpie6}-t_i=
 \simeq\frac{\vec{q}^{\; 2}_i+\xi_i^2m_p^2}{1-\xi_i},\; t_{\pi\pi}=(p_{\pi_1}-p_{\pi_1}^{\prime})^2.
 \end{eqnarray}

\vspace*{-0.7cm}
 The cross-section can be evaluated as follows:
\begin{eqnarray}
&&\frac{d\sigma_{X,{\rm D}\pi {\rm E}}}{d\xi_1 d\xi_2 dt_1 dt_2 d\Phi_X}=\prod\limits_{i=1}^2\left[ \frac{G_{\pi^+pn}^2}{16\pi^2}\frac{-t_i}{(t_i-m_{\pi}^2)^2} F^2(t_i) \xi_i^{1-2\alpha_{\pi}(t_i)} \right]\times\nonumber\\ 
&&\label{bornDpiE}\times\frac{d\sigma_{X,\pi^+ \pi^+}(\xi_1\xi_2 s)}{d\Phi_X} S_2(s/s_0, \{\xi_i\}, \{t_i\}).
\end{eqnarray}
\begin{table}[b!]
\caption{\small\it\label{tab:tot10TeVDpiE}Total $p+p\to n+\pi^++\pi^++n$ cross-sections in the kinematical region $0<|\vec{q}|<0.5\;{\rm GeV}$, $\xi_{min}=10^{-3}<\xi<\xi_{max}$ for two parametrizations from Appendices B (C).}
\begin{center}
\begin{tabular}{|c|c|c|c|c|c|}
\hline
 $\xi_{max}$  &      0.05   & 0.1   & 0.2 & 0.3 & 0.4\\
\hline
 $\sigma_{p+p\to n+\pi^++\pi^++n}$ , $\mu$b  &    0.01 (0.009)   &   0.22 (0.19)  &   3.5 (3.1)  & 12 (11) & 26.8 (24.7) \\
\hline
\end{tabular}
\end{center}
\end{table}
\begin{center}
\vskip -1.6cm
\end{center}
For the $\pi^+\pi^+$ elastic scattering we get
\begin{eqnarray}
&& \label{Udsigmad}\frac{d\sigma_{{\rm el},\; {\rm D}\pi{\rm E}}(s/s_0,\{\xi_i\},\{\vec{q}^{\;2}_i\})}{d\xi_1 d\xi_2 d\vec{q}^{\;2}_1 d\vec{q}^{\;2}_2 dt_{\pi\pi}}=\prod\limits_{i=1}^2 \left[ (m_p^2\xi_i^2+\vec{q}^{\;2}_i)|\Phi_B(\xi_i,\vec{q}^{\;2}_i)|^2 \frac{\xi_i}{(1-\xi_i)^2}\right]\times\nonumber\\
&& \times\; S_2(s/s_0,\{\xi_i\},\{\vec{q}^{\;2}_i\}) \frac{d\sigma_{{\rm el},\;\pi^+\pi^+}(\xi_1\xi_2 s)}{dt_{\pi\pi}},
\end{eqnarray}
\begin{equation}
\label{survivald}S_2=\frac{\sum\limits_{i,j=0,s} \rho_{ij}^2|\bar{\Phi}_{ij}(s/s_0,\{\xi_i\},\{\vec{q}^{\;2}_i\})|^2}{\prod\limits_{i=1}^2\left[(m_p^2\xi_i^2+\vec{q}^{\;2}_i)|\Phi_B(\xi_i,\vec{q}^{\;2}_i)|^2\right]},
\end{equation}
where functions $\rho_{ij}$, $\bar{\Phi}_{ij}$ and $\Phi_B$ are given in the Appendix A. 

\begin{center}
\begin{figure}[t!]
\hskip  2cm \vbox to 4.5cm
{\hbox to 12cm{\epsfxsize=12cm\epsfysize=4.5cm\epsffile{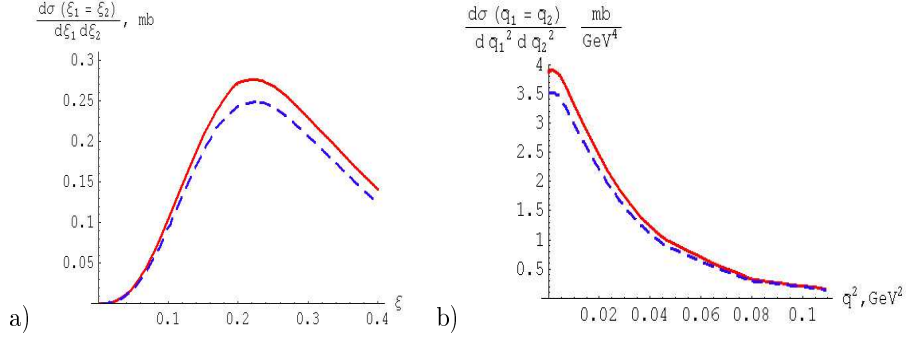}}}
\vskip 0.5cm
\caption{\small\it\label{fig:11}Partially integrated cross-sections of the D$\pi$E process $p+p\to n+\pi^++\pi^++n$ at $\sqrt{s}=10\;{\rm TeV}$ for parametrizations from  Appendices B (solid) and C (dashed): a) $d\sigma/d\xi_1 d\xi_2$ for $\xi_1=\xi_2=\xi$ and $0<|\vec{q}_{1,2}|<0.5\;{\rm GeV}$; b) $d\sigma/d\vec{q}^{\;2}_1 d\vec{q}^{\;2}_2$ for $|\vec{q}_1|=|\vec{q}_2|=|\vec{q}|$ and $10^{-3}<\xi_{1,2}<0.3$.}
\end{figure}
\end{center}

We are now ready to make predictions for high energies. Numerically calculated cross-sections for the exclusive D$\pi$E are shown in Figs.~\ref{fig:11} and listed in Table~\ref{tab:tot10TeVDpiE} for parametrizations from Appendices B and C.

\begin{center}
\vskip -2cm
\end{center}

\section{Extraction of $\pi^+p$ and $\pi^+\pi^+$ cross-sections. The role of absorption. Backgrounds.}

 To extract $\pi^+p$ and $\pi^+\pi^+$ cross-sections from the S$\pi$E and D$\pi$E processes we can use equations~(\ref{born}) and~(\ref{bornDpiE}). Let us rewrite these two equations in the following form
 \begin{eqnarray}
&& \label{eq:pipextract}
 \frac{d\sigma_{X,\pi^+ p}(\xi s)}{d\Phi_X}=\frac{d\sigma_{X,{\rm S}\pi {\rm E}}}{d\xi dt\; d\Phi_X}\frac{E(s/s_0,\xi,t)}{S(s/s_0, \xi, t)},\\
&& \label{eq:pipiextract}
\frac{d\sigma_{X,\pi^+ \pi^+}(\xi_1\xi_2 s)}{d\Phi_X}=\frac{d\sigma_{X,{\rm D}\pi {\rm E}}}{d\xi_1 d\xi_2 dt_1 dt_2 d\Phi_X}\frac{E(s/s_0,\xi_1,t_1)E(s/s_0,\xi_2,t_2)}{S_2(s/s_0, \{\xi_i\}, \{t_i\})},
\end{eqnarray}
where
\begin{equation}
 E(s/s_0,\xi,t)=\frac{(t-m_{\pi}^2)^2}{-t}\frac{16\pi^2}{G_{\pi^+pn}^2F^2(t)\xi^{1-2\alpha_{\pi}(t)}}
\end{equation}

An exact extraction procedure is quite delicate. If we want to extract $\pi^+p$ and $\pi^+\pi^+$ cross-sections in a model independent way, we have to take equations~(\ref{eq:pipextract}),(\ref{eq:pipiextract}) in the limit $t_i\to m_{\pi}^2$. For this limit we should extrapolate the parametrizations of the data on S$\pi$E and D$\pi$E differential cross-sections to the positive $t_i=m_{\pi}^2$, i.e. beyond the physical region. Functions $S$ and $S_2$ are equal to unity for this value of $t_i$, that is why the phenomenological model for these functions is not important. This procedure is actually the Chew-Low extrapolation method~\cite{chewlow},\cite{goebel}.

\begin{center}
\begin{figure}[t!]
\hskip 5cm
 \vbox to 5cm 
{\hbox to 5cm{\epsfxsize=5cm\epsfysize=5cm\epsffile{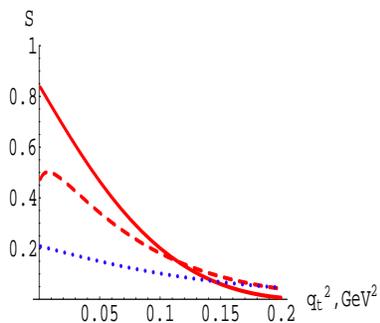}}}
\vskip 0.2cm
\caption{\small\it\label{fig:SSpiE}Function $S(s/s_0,\xi,q_t)$ at $\sqrt{s}=10$~TeV in the physical region of negative $t$ values for three different fixed $\xi$ values: $\xi=0.3$ (dotted), $\xi=0.1$ (dashed) and $\xi=10^{-4}$ (solid). For low $\xi$ and $|\vec{q}|$ function $S$ is close to unity.}
\end{figure}   
\end{center}
\begin{center}
\begin{figure}[t!]
\vbox to 5cm
{\hbox to 15cm{\epsfxsize=15cm\epsfysize=5cm\epsffile{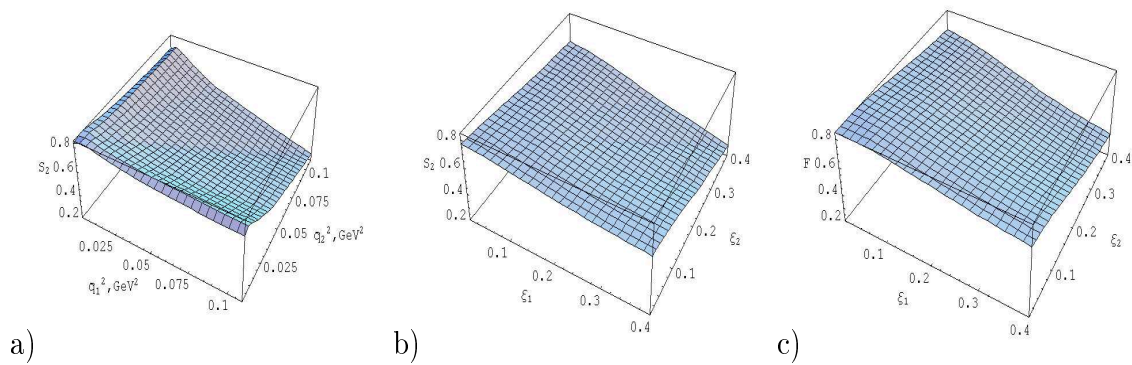}}}
\vskip 0.8cm
\caption{\small\it\label{fig:9}Function $S_2(s/s_0,\xi_{1,2},|\vec{q}_{1,2}|)$ in the physical region of negative $t$ values at  $\sqrt{s}=10$~TeV for: a) for fixed $\xi_{1,2}=0.01$ b) for fixed $|\vec{q}_{1,2}|\sim 0$. c) Function $F(\xi_1,\xi_2)$ at $\sqrt{s}=10$~TeV.}
\end{figure}
\end{center}
\begin{center}
\vskip -2.8cm
\end{center}

Experimentally extrapolation to $m_{\pi}^2$ is rather difficult (see section~\ref{section:exp}), since the errors in $t$ are larger than $m_{\pi}^2$. To get around this problem we extract cross-sections for pions with low virtualities and assume that the values~(\ref{eq:pipextract}),(\ref{eq:pipiextract}) are close to reality. It is clear from the fact that
the main contribution to the cross-section comes 
from the region $|t_i|<0.25\;{\rm GeV}^2$ (see Fig.~\ref{fig:SpiEcs10TeV}b). In this region the dependence
of $\sigma_{\pi^+ p}$ ($\sigma_{\pi^+\pi^+}$) 
on $t$ is assumed to be weak enough.

\begin{center}
\begin{figure}[h!]
\vbox to 5cm
{\hbox to 15cm{\epsfxsize=15cm\epsfysize=5cm\epsffile{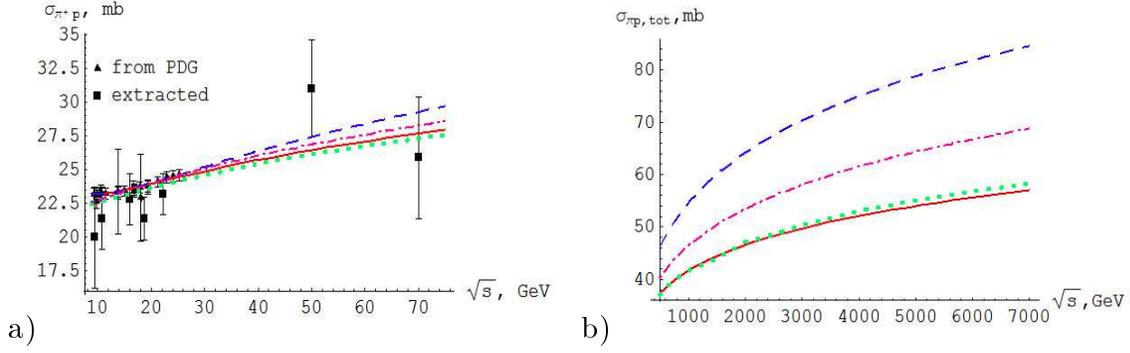}}}
\vskip 0.8cm
\caption{\small\it\label{fig:piptotcs} Total $\pi^+p$ cross-sections versus different parametrizations:~\cite{landshofftot} (solid),\cite{COMPETE} (dashed),\cite{BSW} (dotted) and~\cite{godizov1},\cite{godizov2} (dash-dotted). a) real data from PDG (triangles) up to $\sqrt{s}=25\;{\rm GeV}$ and extracted values (boxes) up to $\sqrt{s}=70\;{\rm GeV}$ (see~\cite{ourneutrontot}); b) total $\pi^+p$ cross-sections in the energy range $0.5\; {\rm TeV}<\sqrt{s}<7\;{\rm TeV}$.}
\end{figure}
\end{center}
\begin{center}
\vskip -2cm
\end{center}
Functions $S$ and $S_2$ are close to unity in the physical region of negative $t$ values (see Figs.~\ref{fig:SSpiE},\ref{fig:9}), and we can estimate errors of the model due to absorptive corrections. It was shown in~\cite{ourneutrontot} that such a model dependent extraction works satisfactory for $\sqrt{s}<70\;{\rm GeV}$. This fact is illustrated in Fig.~\ref{fig:piptotcs}a. All the parametrizations are close to the extracted values and the real data points, but for higher energies (Fig.~\ref{fig:piptotcs}b) the difference between models becomes larger.

\begin{center}
\begin{figure}
\vbox to 5cm
{\hbox to 15cm{\epsfxsize=15cm\epsfysize=5cm\epsffile{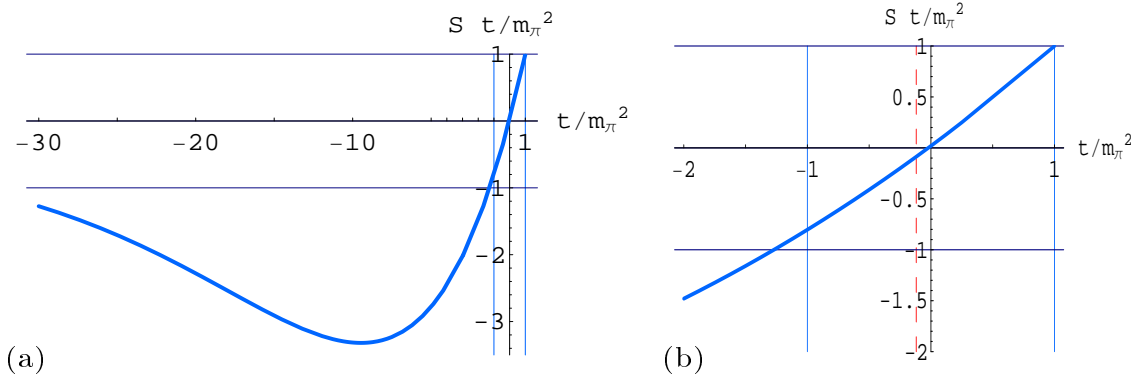}}}
\vskip 0.8cm
\caption{\small\it\label{fig:St0} Function $S(\xi,t)\;t/m_{\pi}^2$ versus $t/m_{\pi}^2$ at fixed $\xi=0.05$. The boundary of the physical region $t_0=-m^2\xi^2/(1-\xi)$ is represented by 
vertical dashed line in b).}
\end{figure}
\end{center}

\begin{center}
\begin{figure}[t!]
\hskip 2cm\vbox to 4cm
{\hbox to 11cm{\epsfxsize=11cm\epsfysize=4cm\epsffile{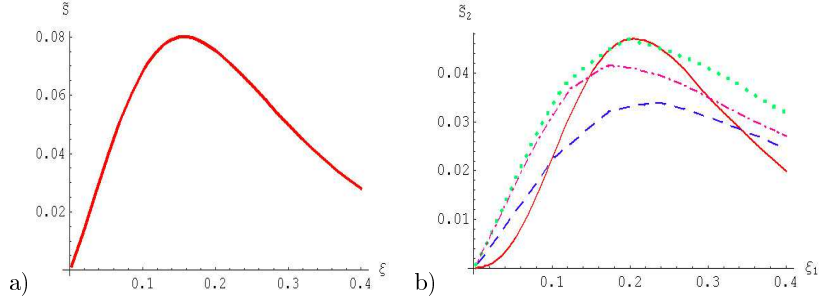}}}
\vskip 0.3cm
\caption{\small\it\label{fig:survint} Values of absorptive corrections integrated with form-factors in the region $0.01\;{\rm GeV}^2<|t_i|<1.2\;{\rm GeV}^2$. a) $\tilde{S}(\xi)$; b) $\tilde{S}_2(\xi_1,\xi_2)$: $\xi_2=\xi_1$ (solid), $\xi_2=0.1$ (dashed), $\xi_2=0.2$ (dotted) and $\xi_2=0.3$ (dash-dotted).}
\end{figure}
\end{center}

To avoid singularities in the extrapolation procedure at $t=0$ and model dependence in $S$ and $S_2$
it is convenient to extrapolate quantities
in the r.h.s of~(\ref{eq:pipextract}) and~(\ref{eq:pipiextract}) multiplied by $S\; t/m_{\pi}^2$ and $S_2\;t_1t_2/m_{\pi}^4$ correspondingly. The behaviour of $S\; t/m_{\pi}^2$ is shown in Fig.~\ref{fig:St0}. It is 
a smooth function of $t$ in the whole region of 
the extrapolation. Practically we will
have $\sigma_{\pi^+p}S\;t/m_{\pi}^2$ as a result of 
the extrapolation, which is equal 
to $\sigma_{\pi^+p}$ at
$t=m_{\pi}^2$.
\begin{center}
\vskip -2cm
\end{center}
The role of absorptive effects (i.e. model dependence of the final result) is significant if we want to extract $\pi p$ and $\pi\pi$ cross-sections from the S$\pi$E and D$\pi$E differential cross-sections integrated in the wide region of $t$ values, where absorption is strong. That is why we need an experimental instrument to measure differential cross-sections for low $t$ values with good resolution. The present design of detectors does not allow $t$ measuremets, it gives only restriction $|t|<\sim 1.2\;{\rm GeV}^2$~\cite{ourneutrontot}. If to 
assume a weak enough $t$-dependence of $\pi p$ and $\pi\pi$ cross-sections, then we could hope to extract these cross-sections (though, with big errors) by the following procedure:
\begin{eqnarray}
 && \label{eq:pipextractINT}
 \sigma_{\pi^+ p}(\xi s)=\frac{\frac{d\sigma_{{\rm S}\pi {\rm E}}}{d\xi}}{\tilde{S}(\xi)},\; \tilde{S}(\xi)=\int\limits_{t_{min}}^{t_{max}}dt \frac{S(s/s_0,\xi,t)}{E(s/s_0,\xi,t)},\\
&& \label{eq:pipiextractINT}
\sigma_{\pi^+ \pi^+}(\xi_1\xi_2 s)=\frac{\frac{d\sigma_{{\rm D}\pi {\rm E}}}{d\xi_1 d\xi_2}}{\tilde{S}_2(\xi_1,\xi_2)},\;\tilde{S}_2(\xi_1,\xi_2)=\int\limits_{t_{min}}^{t_{max}}dt_1 dt_2\frac{S_2(s/s_0, \{\xi_i\}, \{t_i\})}{E(s/s_0,\xi_1,t_1)E(s/s_0,\xi_2,t_2)}.
\end{eqnarray}
Functions $\tilde{S}_2(\xi_1,\xi_2)$ and $\tilde{S}(\xi)$ are depicted in Fig.~\ref{fig:survint}. To  suppress theoretical errors of $\tilde{S}$ and $\tilde{S}_2$  we have to measure total and elastic $pp$ rates at $10$~TeV, since all the models for absorptive corrections are normalized to $pp$ cross-sections. At present we can estimate the theoretical error to be less than 10\% for this method from predicted values of total $pp$ cross-sections in the most popular models (see Fig.~\ref{fig:M1}a).

\begin{center}
\vskip -2cm
\end{center}
The case of the D$\pi$E is more complicated since the function $S_2$ is not factorisable. For low $t_i$ it is approximately equal to
\begin{eqnarray}
&& F(\xi_1,\xi_2)\equiv S_2(s/s_0,\xi_1,\xi_2,0,0)\simeq\nonumber\\
&& \label{survDpiEapprox}\simeq \left( \sqrt{S(s/s_0,\xi_1,0)}+\sqrt{S(s/s_0,\xi_2,0)}- \sqrt{S(s/s_0,\xi_1,0)S(s/s_0,\xi_1,0)} \right)^2,
\end{eqnarray}
which is clear from Figs.~\ref{fig:9}b,c.

To estimate total absorptive effect we have to take the ratio
\begin{eqnarray}
\label{eq:totabsorb}
&& S_{\rm tot}=\int\limits_{\Omega^{\prime}}\frac{d\sigma}{d\Phi}\left/ \int\limits_{\Omega^{\prime}}\frac{d\sigma_0}{d\Phi}\right.,\\
&& \Omega^{\prime}:\; \xi_{\rm min}=10^{-3}<\xi_i<\xi_{\rm max},\; 0.01\;{\rm GeV}^2<|t|<0.5\;{\rm GeV}^2.
\end{eqnarray}
$\Phi$ is the phase space for the S$\pi$E (D$\pi$E), and $d\sigma_0/d\Phi$ is the cross-section without absorptive corrections (i.e. for $S\equiv S_2\equiv 1$). The results are listed in Table~\ref{tab:Stot}.

\begin{table}
\caption{\small\it\label{tab:Stot}Total absorptive corrections for exclusive S$\pi$E and D$\pi$E in the kinematical region $0.01\;{\rm GeV}^2<|t_i|<0.5\;{\rm GeV}^2$,  $\xi_{min}=10^{-3}<\xi_i<\xi_{max}$ for parametrizations from Appendices B,C.}
\begin{center}
\begin{tabular}{|c|c|c|c|c|c|}
\hline
 $\xi_{max}$  &      0.05   & 0.1   & 0.2 & 0.3 & 0.4\\
\hline
 $S_{{\rm tot},\; {\rm S}\pi{\rm E}}$  &    $0.39$   &   $0.34$  &   $0.27$  & $0.22$ & $0.18$\\
\hline
 $S_{{\rm tot},\; {\rm D}\pi{\rm E}}$  &    $0.47$   &   $0.41$  &   $0.33$  & $0.26$ & $0.25$\\
\hline
\end{tabular}
\end{center}
\end{table}

\begin{center}
\begin{figure}[b!]
\hskip  2cm \vbox to 6cm
{\hbox to 10cm{\epsfxsize=10cm\epsfysize=6cm\epsffile{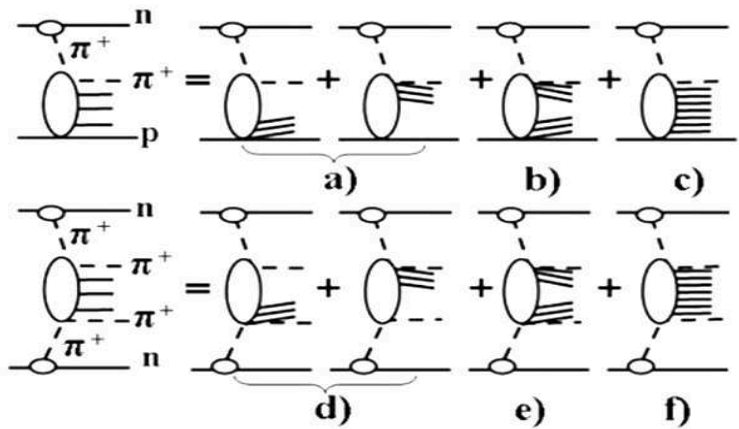}}}
\vskip 0.2cm
\caption{\small\it\label{fig:piexpand} Inclusive S$\pi$E (a,b,c) and D$\pi$E (d,e,f) backgrounds. a) single dissociation, b) double dissociation, c) minimum bias in the $\pi^+p$ channel, and d) single dissociation, e) double dissociation, f) minimum bias in the $\pi^+\pi^+$ channel.}
\end{figure}
\end{center}

 If we could measure momenta of all the final particles, we would have only exclusive backgrounds for our S$\pi$E and D$\pi$E signal processes with elastic $\pi^+ p$ (Fig.~\ref{fig:1}a) and $\pi^+\pi^+$ (Fig.~\ref{fig:1}c) scattering. The single low mass dissociative background for the exclusive S$\pi$E is depicted in the Fig.~\ref{fig:1}b. For the exclusive D$\pi$E we have two exclusive backgrounds of Figs.~\ref{fig:1}d,e. In a real experiment we can detect only one or two particles in the final state, and we have to take into account all the inclusive backgrounds: single and double dissociation, central diffraction, minimum bias with neutrons production, inclusive single and double charge exchanges with $\pi^+$ (see Fig.~\ref{fig:piexpand}) and also $\rho$, $a_2$ exchanges (for more exact estimations of inclusive backgrounds see Section~\ref{section:exp}). 

\begin{center}
\vskip -2cm
\end{center}

\section{Experimental possibilities}
\label{section:exp}

In this chapter we analyse CMS~\cite{CMS}  capabilities to measure
elastic $\pi^+ p$ and $\pi^+ \pi^+$ scattering at 10 TeV, c.m. energy of LHC protons in the first runs.  The CMS Zero Degree Calorimeters ,ZDCs, ~\cite{ZDC,Grachov:2008qg} can measure leading neutrons in the exclusive  S$\pi$E, $pp \to n p \pi^+$ (Fig.~\ref{fig:1}a), and 
 D$\pi$E, $pp \to n \pi^+ \pi^+ n$  (Fig.~\ref{fig:1}c), processes\footnote{
Further, exclusive S$\pi$E (D$\pi$E) elastic events, i.e.  S$\pi$E (D$\pi$E) with   $\pi^+_{virt}p$ ($\pi^+_{virt}\pi^+_{virt}$) scattering elasticaly, will be designated as
S$\pi$E$_{elastic}$ (D$\pi$E$_{elastic}$) for brevity.}. 
The ZDCs are located  between the two beam pipes at ~140 m 
on each side of the interaction point. They are able to measure the energy of forward 
neutral particles in the pseudorapidity region $|\eta|>8.5$. 
\begin{figure}[b!]
\begin{center} 
\hskip  0.5cm \vbox to 4.5cm
{\hbox to 13cm{\epsfxsize=13cm\epsfysize=4.5cm\epsffile{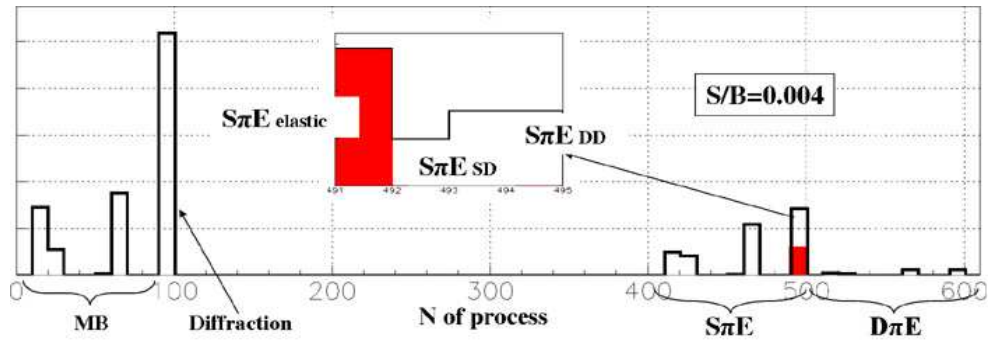}}}
\caption{\small\it\label{exp:f2}Events ratio for S$\pi$E$_{elastic}$ (shadowed) and background.}
\end{center} 
\end{figure}
 S$\pi$E  and D$\pi$E events  have been generated
in the framework of the simulation package EDDE~\cite{EDDE}. The kinematics of the S$\pi$E  and D$\pi$E proceses are defined by $\xi_n$ and $t_n$ of the leading neutron. The $p\pi^+_{virt}n$ vertex is generated on the basis of the model described in  Ref.~\cite{ourneutrontot}. For the simulation of 
$\pi^+_{virt}p$ and $\pi^+_{virt}\pi^+_{virt}$ elastic scattering  PYTHIA 6.420~\cite{pythia} has been used.  Obviously, $\pi^+_{virt}$ and $p$  ($\pi^+_{virt}$ and $\pi^+_{virt}$) can interact inelasticaly and diffractively. Then, in the diffractive interaction of  $\pi^+_{virt}$ and $p$, either the  $\pi^+_{virt}$  or the $p$, or both of them,  can  dissociate. All of these processes have been studied as backgrounds, as well as minimum bias and diffractive $pp$ events. Diagrams for some of the background processes are shown in the Fig.~\ref{fig:piexpand}. Signal and background have been generated by EDDE v.3.0.0 and PYTHIA 6.420.  The S$\pi$E cross section, including all types of $\pi^+_{virt}p$ interactions, is estimated to be about 2.6 mb at 10 TeV and  $\xi_n<0.4$~\cite{ourneutrontot}.  Corresponding cross sections for the signal and backgrounds are listed below.

\begin{itemize}
\item[] Signal:
\begin{itemize}
\item[-]
S$\pi$E$_{elastic}$ :   $\sigma_{pp \to n p \pi^+} = 0.33$ mb. 
\end{itemize}
\item[] Backgrounds from $pp$ and inelastic S$\pi$E events: 
\begin{itemize}
\item[-]
minimum bias events: $\sigma_{pp \to X} = 50$ mb;
\item[-]
single diffractive dissociation: $\sigma_{pp \to pX} = 14$ mb ;
\item[-]
double diffractive dissociation: $\sigma_{pp \to XY} = 9.7$ mb;
\item[-]
S$\pi$E, minimum bias in the $\pi^+_{virt}p$ channel, Fig.~\ref{fig:piexpand}c: 
$\sigma_{pp \to n X} = 1.54$ mb;
\item[-]
S$\pi$E, single diffraction in the $\pi^+_{virt}p$ channel with proton dissociation, 
Fig.~\ref{fig:piexpand}a: $\sigma_{pp \to n \pi^+ X} = 0.23$ mb;
\item[-]
S$\pi$E, single diffraction in the $\pi^+_{virt}p$ channel with  $\pi+$ dissociation, 
Fig.~\ref{fig:piexpand}a: $\sigma_{pp \to n p X} = 0.20$ mb;
\item[-]
S$\pi$E,  double diffraction in the $\pi^+_{virt}p$ channel,  
Fig.~\ref{fig:piexpand}b:  $\sigma_{pp \to nXY} = 0.27$ mb.
\end{itemize}
\end{itemize}

Fig.~\ref{exp:f2} shows the ratio of events for S$\pi$E$_{elastic}$ (shadowed) and background processes. On the picture minimum bias processes have numbers less than 90 and diffractive  processes have numbers 92, 93 and 94, according to PYTHIA's definition. All  S$\pi$E processes  are placed between numbers 400 and 500 and D$\pi$E processes are in the region from 500 to 600 in our generation. Signal processes, S$\pi$E$_{elastic}$ and D$\pi$E$_{elastic}$, have numbers 491 and 591 respectively. S$\pi$E$_{elastic}$
events contribute $\sim$0.4\%  to the total cross sections, D$\pi$E$_{elastic}$ events is around 0.025\% only.

D$\pi$E events have been simulated by the same method, as  S$\pi$E, using EDDE v.3.0.0 and PYTHIA 6.420.  D$\pi$E cross section, including all types of $\pi^+_{virt}\pi^+_{virt}$ interactions, is estimated to be about 200 $\mu$b at 10 TeV and  $\xi_{n1,n2}<0.4$~\cite{ourneutrontot}.  Corresponding cross sections for the signal and inelastic D$\pi$E backgrounds are listed below.
\begin{itemize}
\item[] Signal:
\begin{itemize}
\item[-]
D$\pi$E$_{elastic}$:   $\sigma_{pp \to n \pi^+ \pi^+ n} = 24$ $\mu$b. 
\end{itemize}
\item[] Backgrounds from inelastic D$\pi$E events:
\begin{itemize}
\item[-]
D$\pi$E, minimum bias in the $\pi^+_{virt}\pi^+_{virt}$ channel, 
Fig.~\ref{fig:piexpand}(f): $\sigma_{pp \to n X n} = 124$  $\mu$b;
\item[-]
D$\pi$E, single diffraction in the $\pi^+_{virt}\pi^+_{virt}$ channel, 
Fig.~\ref{fig:piexpand}d: $\sigma_{pp \to n \pi^+ X n} = 30$  $\mu$b;
\item[-]
D$\pi$E,  double diffraction in the $\pi^+_{virt}\pi^+_{virt}$ channel,  
Fig.~\ref{fig:piexpand}(e):  $\sigma_{pp \to nXYn} = 22$~$\mu$b.
\end{itemize}
\end{itemize}
Inelastic and diffractive $pp$ interactions produce background events 
for  D$\pi$E$_{elastic}$, as well as for S$\pi$E$_{elastic}$.
Moreover, S$\pi$E elastic and inelastic processes produce strong backgrounds for   
D$\pi$E$_{elastic}$, in addition to the inelastic $pp$ and D$\pi$E.
And, on the contrary,  D$\pi$E elastic and inelastic events can imitate S$\pi$E$_{elastic}$.
\begin{figure}[b!]
\begin{center} 
\hskip 0.1cm\vbox to 4.5cm
{\hbox to 13cm{\epsfxsize=13cm\epsfysize=4.5cm\epsffile{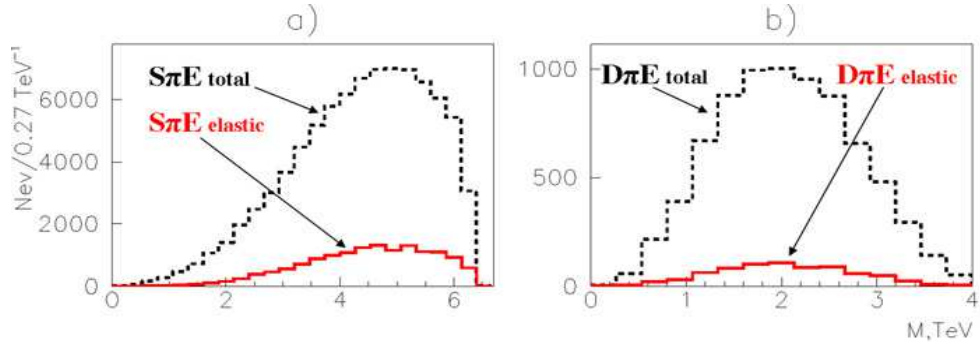}}}
\caption{\small\it \label{exp:f1}
(a)  S$\pi$E elastic  (solid) and total (dotted) events distribution
versus the ($\pi^+_{virt}p$) invariant mass;
(b)  D$\pi$E elastic  (solid) and total (dotted) events distribution
versus the ($\pi^+_{virt}\pi^+_{virt}$) invariant mass.}
\end{center} 
\end{figure}  
In our simulation, cross sections for the 
S$\pi$E$_{elastic}$  and  D$\pi$E$_{elastic}$ signals, depend
on $\pi^+_{virt}p$ and  $\pi^+_{virt}\pi^+_{virt}$ elastic scattering models 
integrated to the PYTHIA 6.420. It is interesting to note  that values obtained
 for these cross sections are very close to those ones which can be calculated in the BSW~\cite{BSW} and GP~\cite{godizov1, godizov2} parametrizations, see Table~\ref{tab:tot10TeVnum} and~\ref{tab:tot10TeVDpiE}. The ratio between the S$\pi$E elastic (signal) and inelastic (part of background) events are presented in Fig.~\ref{exp:f1}a. Fig.~\ref{exp:f1}b shows the same for D$\pi$E.

\begin{figure}[b!]
\begin{center} 
\hskip 0.5cm\vbox to 4.5cm
{\hbox to 13cm{\epsfxsize=13cm\epsfysize=4.5cm\epsffile{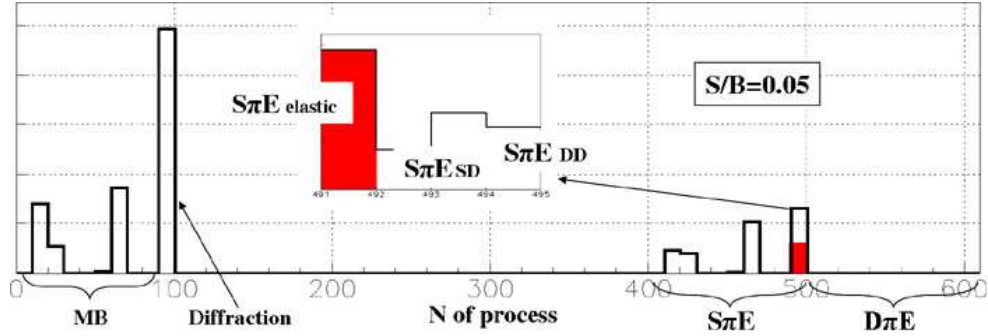}}}
\caption{\small\it\label{exp:f3}
The ratio of events for the signal S$\pi$E$_{elastic}$ (shadowed) and background  processes after the selection~(\ref{exp:sel1}). 
}
\end{center} 
\end{figure} 

As in paper~\cite{ourneutrontot}, for S$\pi$E
selections we choose events with neutrons in the forward or backward 
ZDC and with the absence of neutrons in the opposite one:
\begin{equation}
\label{exp:sel1}
\left[
\begin{array}{l} 
\rm N_n^f>0\quad  \& \quad N_n^b=0\\
\rm N_n^b>0\quad  \& \quad N_n^f=0.
\end{array}
\right.
\end{equation}
For the D$\pi$E, we selected events with neutrons in both the forward and 
backward ZDCs:
\begin{equation}
\label{exp:sel2}
\begin{array}{l} 
\rm N_n^f>0\quad  \& \quad N_n^b>0.
\end{array}
\end{equation} 

\begin{figure}[b!]
\begin{center} 
\hskip 0.1cm\vbox to 4.5cm
{\hbox to 13cm{\epsfxsize=13cm\epsfysize=4.5cm\epsffile{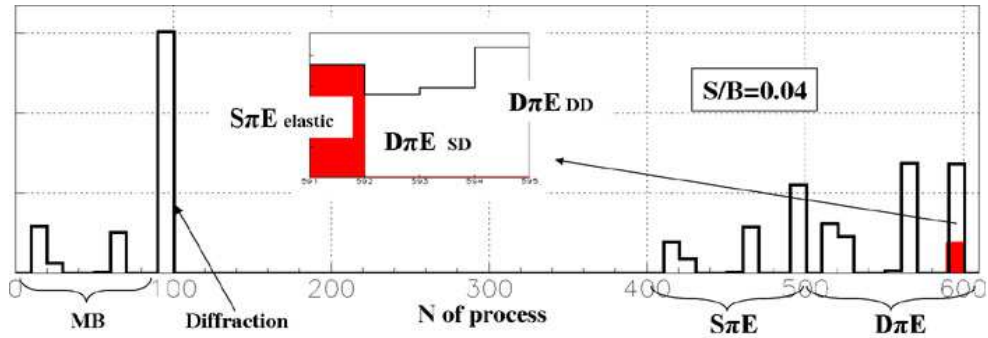}}}
\caption{\small\it\label{exp:f4}
The ratio of events for the signal D$\pi$E$_{elastic}$ (shadowed) and background  processes after the selection~(\ref{exp:sel2}).
}
\end{center} 
\end{figure}

Here, $N_n^f$ ($N_n^b$) is the number of neutrons hitting 
the forward (backward) ZDC.
Such selections suppress the background for S$\pi$E$_{elastic}$ (D$\pi$E$_{elastic}$) events by a factor 14 (160). The signal to background ratio becomes equal to 0.05 for the  
S$\pi$E$_{elastic}$ and 0.04 for the D$\pi$E$_{elastic}$.
Fig.~\ref{exp:f3} shows the ratio of events for S$\pi$E$_{elastic}$ (shadowed) and background processes after the selection~(\ref{exp:sel1}). The same picture is plotted for the D$\pi$E$_{elastic}$ signal and background after the  selection~(\ref{exp:sel2}), Fig.~\ref{exp:f4}. 
\begin{figure}[t!]
\begin{center} 
 \hskip 0.1cm\vbox to 4.5cm
{\hbox to 13cm{\epsfxsize=13cm\epsfysize=4.5cm\epsffile{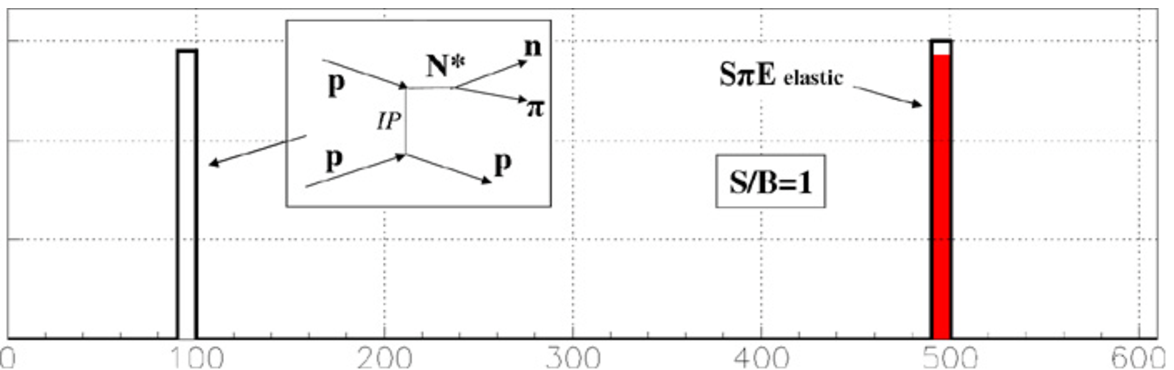}}}
\caption{\small\it\label{exp:f5}
The ratio of events for the signal S$\pi$E$_{elastic}$ (shadowed) and background  processes after selections~(\ref{exp:sel1}) {\rm \&}~(\ref{exp:sel3}).
}
\end{center} 
\end{figure} 
\begin{figure}[t!]
\begin{center} 
\vbox to 5cm
{\hbox to 15cm{\epsfxsize=15cm\epsfysize=5cm\epsffile{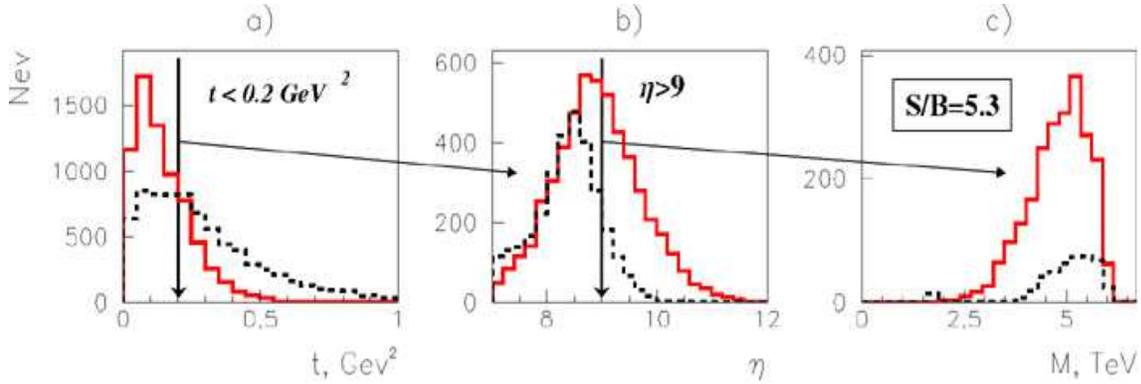}}}
\caption{\small\it\label{exp:f7}
a) $t$ of the leading neutron for the signal S$\pi$E$_{elastic}$ (solid) and  
background (dotted)  after selections~(\ref{exp:sel1}) {\rm \&}~(\ref{exp:sel3});
b) $\eta$ of the $\pi^+$  for the signal S$\pi$E$_{elastic}$ (solid) and background (dotted) 
after selections~(\ref{exp:sel1}) {\rm \&}~(\ref{exp:sel3}) {\rm \&} $t_n<0.2\;{\rm GeV}^2$;
c) the ($\pi^+p$) mass distribution for the signal S$\pi$E$_{elastic}$ (solid) and  background (dotted) after selections~(\ref{exp:sel1}) {\rm \&}~(\ref{exp:sel3})
{\rm \&} $t_n<0.2\;{\rm GeV}^2$ {\rm \&}  $\eta_{\pi^+}>8.5$.
}
\end{center} 
\end{figure}
 
The signal S$\pi$E$_{elastic}$ ($pp \to n \pi^+ p$) event has neutron, proton
and $\pi^+$ in the final state.  Apart from neutrons, which can be detected 
by ZDC, two other particles move out of the CMS acceptance. The proton, scattered elasticaly, should move inside the beam pipe. The $\pi^+$ meson should fly in the same direction as the neutron and it is deflected at a small angle too (see Fig.~\ref{exp:f7}b).
Thus,  though it looks as a paradox, we should demand an absence of a signal
in the CMS detectors, except of  the one of  ZDCs, for the S$\pi$E$_{elastic}$ trigger. For example, we could select events with zero signal in the CMS calorimeters:
\begin{equation}
\label{exp:sel3}
\begin{cases}{
\begin{array}{l} 
\rm N_{BARREL}=0\\
\rm N_{ENDCAP}=0\\
\rm N_{HF}=0\\
\rm N_{CASTOR}=0.
\end{array}
}\end{cases}
\end{equation} 
Fig.~\ref{exp:f5} shows  the efficiency of such selection for the S$\pi$E$_{elastic}$.
The signal to background ratio becomes equal to 1, i.e. we have achieved improvement 18 times better in comparison with the previous selection~(\ref{exp:sel1}). Detailed study of the rest of backgrounds (left bin on the Fig.~\ref{exp:f5}) has shown that it contains processes of the
single diffractive dissociation, $pp \to p N^* \to p n \pi^+$, where the Pomeron exchange leads to the proton excitation $N^*$ and its subsequent decay to the $\pi^+$ meson and neutron (see the diagram on the Fig.~\ref{exp:f5}). This reaction can imitate the S$\pi$E$_{elastic}$ process as well. However, the further careful study of both reactions has shown some difference in their kinematics. Thus,  $t$ distributions  of neutrons have different slope parameters. It could improve the signal/background ratio up to the value 1.7 by the selection $|t_n|<0.2\;{\rm GeV}^2$, Fig.~\ref{exp:f7}a. 

\begin{figure}[t!]
\begin{center} 
\hskip 0.1cm \vbox to 6cm
{\hbox to 15cm{\epsfxsize=15cm\epsfysize=6cm\epsffile{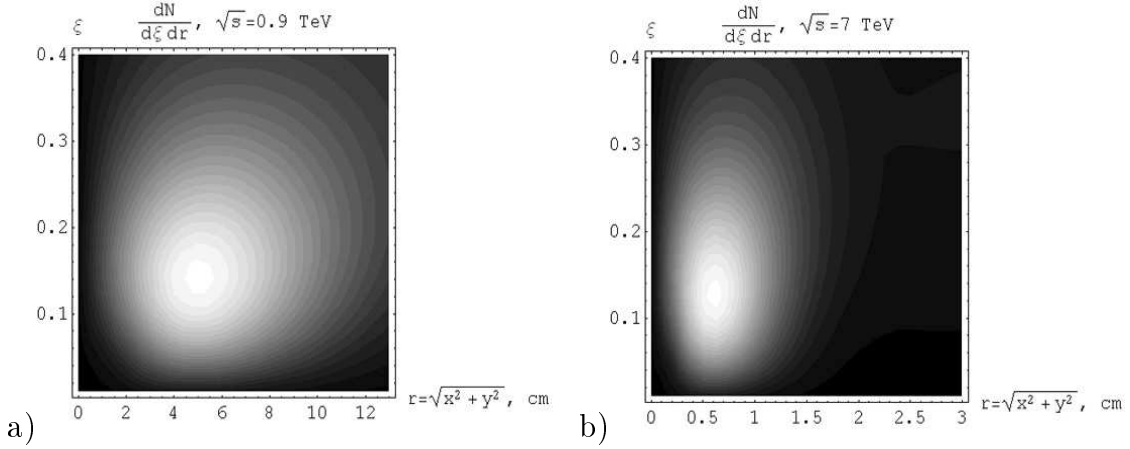}}}
\vskip 0.2cm
\caption{\label{dNdXidr}
$dN/d\xi dr$ for a) $\sqrt{s} = 900$~GeV and b) $7$~TeV}
\end{center} 
\end{figure} 
\begin{figure}[tb!]
\begin{center} 
\hskip 0.1cm \vbox to 6cm
{\hbox to 15cm{\epsfxsize=15cm\epsfysize=6cm\epsffile{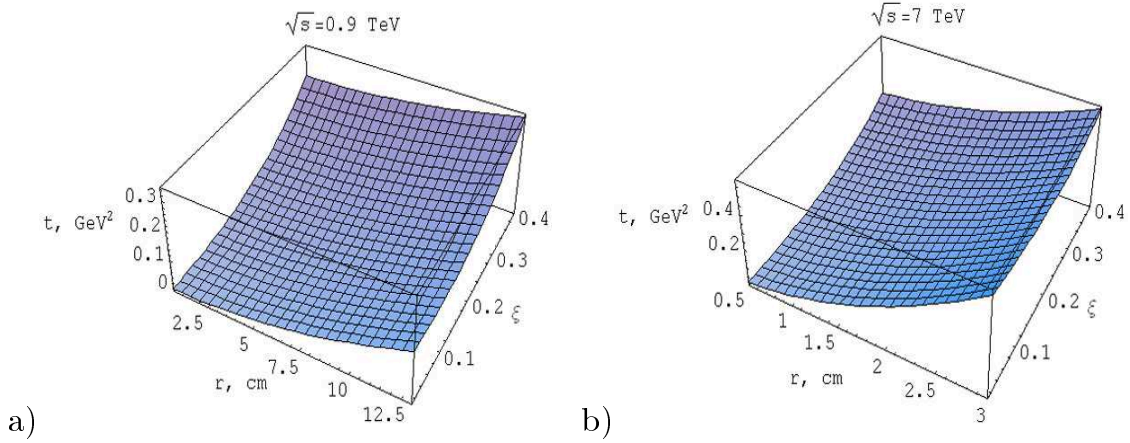}}}
\vskip 0.2cm
\caption{\label{TvsRXi} The $t$ value for neutrons as a function of distance from the collision axis and $\xi$
for a) $\sqrt{s} = 900$~GeV and b) $7$~TeV}
\end{center} 
\end{figure} 
Each of the CMS ZDCs consist of two sections, an electromagnetic, EM,  part  for measuring photons, $\pi_0$, $\eta$s etc and a hadronic part designed to measure neutral hadrons such as neutrons and $\lambda$s~\cite{Grachov:2008qg}. The energy resolution of the detector for hadrons is 
  $138\%/\sqrt{E} + 13\%$. The electromagnetic part is divided into strips that run in the vertical direction. These strips can be used to  measure the horizontal position of the particle's impact point with a resolution of about 0.4cm. The hadronic part that has is divided into 4 depth segments but has no transverse segmentation.   About 1/3$^{rd}$ of the time neutrons will start to shower in the electromagnetic part and for these neutrons we can extract some position information.  The ZDCs  can also be used to select events in the CMS level one trigger. 

 The geometrical acceptance  of the calorimeter is $\pm 4.4$ cm horizontally and $\pm 5.0$ vertically. For CMS the LHC beams cross in the horizontal plane and so the nominal position of the zero degree point will vary depending on the crossing angle. For example if the crossing angle of the beam is 140$\mu$radians the zero degree point will be at x=+2cm.  Given the energy and position resolution of the detector it may be possible to make a rough measurement of the angular distribution of the neutrons.  

\begin{figure}
\begin{center} 
\hskip 0.1cm \vbox to 6cm
{\hbox to 15cm{\epsfxsize=15cm\epsfysize=6cm\epsffile{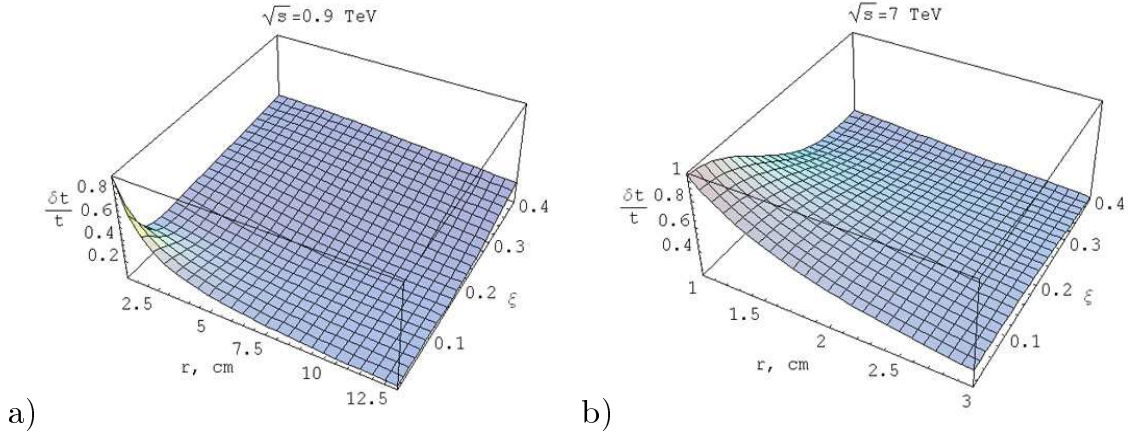}}}
\vskip 0.2cm
\caption{\label{Tresolution} The relative resolution for  $t$, i.e. $\delta t/t$  value for neutrons as a function of distance from the collision axis and $\xi$
for a) $\sqrt{s} = 900$~GeV and b) $7$~TeV. We have assumed that $\delta r$ = 0.5cm, distance from the detector to the interaction point is $140$~m and $\delta\xi/\xi\simeq 0.14$}
\end{center} 
\end{figure}

The only independent measurements that the ZDCs can make of the neutrons are the energy loss $\xi$ and the distance from the collision axis $r=\sqrt{x^2+y^2}$.  
Figures \ref{dNdXidr}, \ref{TvsRXi} and \ref{Tresolution} show the distribution of neutrons, their $t$ value and the resolution of $t$ versus $\xi$ and r for a) $\sqrt{s} = 900$~GeV and b) $7$~TeV.  As the energy increases the radial distribution tends to shrink towards $r=0$. Given the current limited position resolution of the ZDC it may be possible to gain some information about the $t$ distribution at $\sqrt{s} = 900$~GeV.  However for multi TeV energies it will probably be necessary to upgrade the detector. 
Figure ~\ref{exp:f7}(b) shows that the $\pi^+$ mesons are deflected a bit stronger for the signal than for a background,  If we could have counters for charge particles in the pseudorapidity region $\eta>9$, it would allow us to improve the signal/background ratio up to 5 and higher, Fig.~\ref{exp:f7}c.
The possibility of such counters installation along the LHC beam on both sides of the CMS was studied in the Ref.~\cite{FSC}.  Set of FSCs, placed
at distances from 60 to 140 m from the interaction point, could cover the pseudorapidity region from 8 to 11. They could  register particle showers induced by the primary $\pi^+$ with high efficency, up to 70\%.
 Unfortunately, in the present setup of the forward CMS detectors there are no forward counters and  present design of the ZDC does not allow to measure $t$ of the leading neutron. So, this is a task for the future.

\begin{figure}[tb!]
\begin{center} 
\hskip 0.1cm \vbox to 4.5cm
{\hbox to 13cm{\epsfxsize=13cm\epsfysize=4.5cm\epsffile{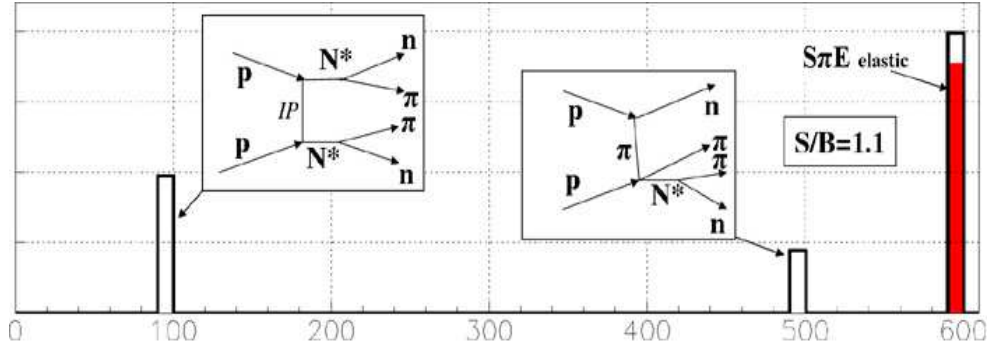}}}
\vskip 0.2cm
\caption{\small\it\label{exp:f6}
The ratio of events for the signal D$\pi$E$_{elastic}$ (shadowed) and background  processes after selections~(\ref{exp:sel2}) {\rm \&}~(\ref{exp:sel3}).
}
\end{center} 
\end{figure}
\begin{figure}[tb!]
\begin{center} 
\hskip 0.1cm \vbox to 4.5cm
{\hbox to 13cm{\epsfxsize=13cm\epsfysize=4.5cm\epsffile{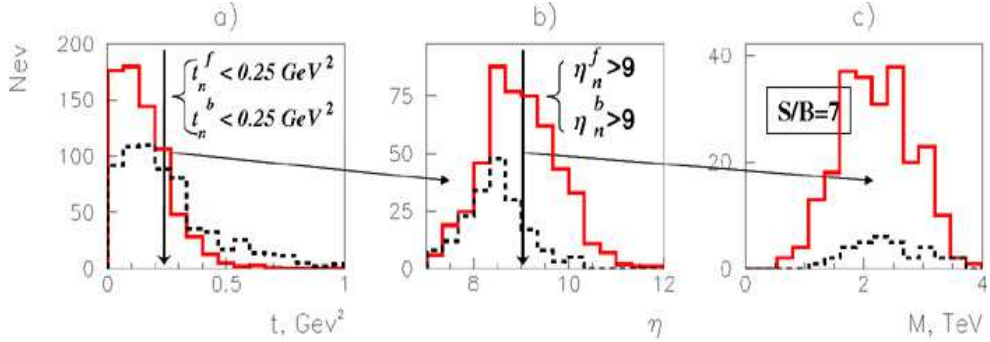}}}
\caption{\small\it\label{exp:f8}
a) $t$ of the leading neutrons for the signal D$\pi$E$_{elastic}$ (solid) and  
background (dotted)  after selections~(\ref{exp:sel2}) {\rm \&}~(\ref{exp:sel3});
b) $\eta$ of the $\pi^+$ mesons  for the signal D$\pi$E$_{elastic}$ (solid) and background (dotted) 
after selections~(\ref{exp:sel2}) {\rm \&}~(\ref{exp:sel3}) {\rm \&}~(\ref{exp:sel4}) ;
c) the ($\pi^+\pi^+$) mass distribution for the signal D$\pi$E$_{elastic}$ (solid) and  background (dotted) after selections~(\ref{exp:sel2}) {\rm \&}~(\ref{exp:sel3}) {\rm \&}~(\ref{exp:sel4})  {\rm \&}~(\ref{exp:sel5}). 
}
\end{center} 
\end{figure}
 Selections~(\ref{exp:sel3}), applied to the D$\pi$E$_{elastic}$, improve the  
signal/background ratio up to 1.1 (see Fig.~\ref{exp:f6}).  The rest of the background comes from the the double diffractive dissociation $pp \to  N^* N^*  \to n \pi^+ \pi^+ n$
(the left diagram on the  Fig.~\ref{exp:f6}) and  from the S$\pi$E events produced by single diffraction in the $\pi^+p$  channel with the subsequent decay of the excited protons to $\pi^+$ and neutron (the right diagram on the  Fig.~\ref{exp:f6}).   
$t$ distributions for signal and background are different, as it is shown on the Fig.~\ref{exp:f8}a.
As for the S$\pi$E, we could improve the signal/background ratio up to $\sim2$ by selections:
\begin{equation}
\label{exp:sel4}
\begin{cases}{
\begin{array}{l} 
 |t_n^f|<0.25\; {\rm GeV}^2,\\
 |t_n^b|<0.25\; {\rm GeV}^2,
\end{array}
}\end{cases}
\end{equation} 
using~(\ref{exp:sel4}) in a combination with~(\ref{exp:sel2}) and~(\ref{exp:sel3}). 
For further improvements of data we could use different deviation of  $\pi$ meson from the beam axis for
signal and background,  Fig.~\ref{exp:f8}b. Selections
\begin{equation}
\label{exp:sel5}
\begin{cases}{
\begin{array}{l} 
 \eta_n^f>9,\\
 \eta_n^b>9,
\end{array}
}\end{cases}
\end{equation} 
give a data sample with a signal/background ratio $\sim7$, see  Fig.~\ref{exp:f8}c.

\section{Discussions and conclusions}

In conclusion, our study of S$\pi$E$_{elastic}$  and D$\pi$E$_{elastic}$ processes 
shows that with present setup of the forward detectors we could expect observation of S$\pi$E$_{elastic}$ and D$\pi$E$_{elastic}$ events mixing with background in the  proportion $\sim$1:1.
Rough estimations on the generator level shows that we could observe $\sim10^8$  S$\pi$E$_{elastic}$ events distributed in the mass region from 1 to 6 TeV
and  $\sim10^7$  D$\pi$E$_{elastic}$ events distributed in the mass region from 0.5 to 4 TeV at the integrated luminosity 1 pb$^{-1}$.  As it was said they will be mixed with approximately the same amount of background events. 
Improvement of  the data purity demands a considerable modernisation of the forward detectors.
Some modification of the ZDC is required to measure the $t$ of the leading neutrons. 
It would be very useful to install  forward shower counters FSCs along the beam at  distancies from 60 to 140 m  for the detection of  elastic scattered  $\pi$-mesons  in the region  $\eta > 8$ would improve the measurements significantly.  Realisation of such modifications  is beyond this article. 

 Theoretically, it is very interesting to have both elastic and total cross-sections of
 $\pi p$ and $\pi\pi$ scattering. At present we could only use the extraction 
 procedure for $t$-integrated S$\pi$E and D$\pi$E cross-sections~(\ref{eq:pipextractINT}),(\ref{eq:pipiextractINT}) which is far from the ideal one~(\ref{eq:pipextract}),(\ref{eq:pipiextract}). Rough estimations give the model error about 10\%.  
The main part of this error comes from the uncertainties in the absorptive 
corrections which are normalized to $pp$ total and elastic cross sections. Measurements of the $pp$ total cross section, which would be done by the TOTEM experiment at LHC, can improve the precision of our model-dependent extraction procedure  significantly. We would like to stress again, that model-independent extraction procedure for  $\pi p$ and $\pi\pi$ total and elastic cross sections makes precision measurements of $t$ of the leading neutron at small angles mandatory.

In spite of all the difficulties, proposed tasks are of exceptional importance, and we hope that they will push modernisation of the forward detectors for future precise measurements.

\section*{Appendix A}

 Here we define functions for the calculation of absorptive corrections for the S$\pi$E and the D$\pi$E processes. For the S$\pi$E we have:
\begin{eqnarray}
&& \Phi_B(\xi,\vec{q}^{\;2})=\frac{N(\xi)}{2\pi}\left(
\frac{1}{\vec{q}^{\;2}+\epsilon^2}+\imath \frac{\pi\alpha_{\pi}^{\prime}}{2(1-\xi)}
\right)\exp(-\beta^2\vec{q}^{\;2})\simeq\nonumber\\
&& \simeq \frac{N(\xi)}{2\pi}
\frac{1}{\vec{q}^{\;2}+\epsilon^2}
\frac{1}{1+\beta^2\vec{q}^{\;2}},\; \vec{q}\to 0,\\
&& N(\xi)=(1-\xi)\frac{G_{\pi^+pn}}{2}\xi^{\frac{\alpha_{\pi}^{\prime}\epsilon^2}{1-\xi}}\exp\left[ -b\frac{m_p^2\xi^2}{1-\xi}\right],\\
&&\beta^2=\frac{b+\alpha_{\pi}^{\prime}\ln\frac{1}{\xi}}{1-\xi},\; \epsilon^2=m_p^2\xi^2+m_{\pi}^2(1-\xi),\\
&& \Phi_0=\frac{N(\xi)}{2\pi}\int\limits_0^{\infty} db\; \Theta_0(b,\xi,|\vec{q}|)V(b),\\
&& |\vec{q}|\Phi_s=\frac{N(\xi)}{2\pi}\int\limits_0^{\infty} db\;\Theta_s(b,\xi,|\vec{q}|) V(b),\\
&& \Theta_0(b,\xi,|\vec{q}|)=\frac{b\; J_0(b|\vec{q}|)\left(K_0(\epsilon\;b)-K_0\left(\frac{b}{\beta}\right)\right)}{1-\beta^2\epsilon^2},\\
&& \Theta_s(b,\xi,|\vec{q}|)=\frac{b\; J_1(b|\vec{q}|)\left(\epsilon\; K_1(\epsilon\; b)-\frac{1}{\beta}K_1\left( \frac{b}{\beta}\right)\right)}{1-\beta^2\epsilon^2},
\end{eqnarray}
\begin{eqnarray}
&& V(b)=\exp\left( -\Omega_{el}(s/s_0,b)\right),\\
&& \label{U3}\Omega_{el}=\sum\limits_{i=1}^3 \Omega_i,\; \Omega_i=\frac{2c_i}{16\pi B_i}\left(\frac{s}{s_0}{\rm e}^{-\imath\frac{\pi}{2}} \right)^{\alpha_{IP_i}(0)-1}\exp\left[ -\frac{b^2}{4B_i}\right],\\
&& \label{U4}B_i=\alpha^{\prime}_{IP_i}\ln\left(\frac{s}{s_0}{\rm e}^{-\imath\frac{\pi}{2}} \right)+\frac{r_i^2}{4},
\end{eqnarray}
the values of parameters can be found in~(\ref{3IPtrajectories}) and in Table~\ref{tab:3IP}.

 For the D$\pi$E process we can write the following expressions:
\begin{eqnarray}
&& \bar{\Phi}_{ij}=\frac{N(\xi_1)N(\xi_2)}{(2\pi)^2}\int\limits_0^{\infty} db_1 db_2 \Theta_i(b_1,\xi_1,|\vec{q}_1|)\Theta_j(b_2,\xi_2,|\vec{q}_2|) I_{\phi}(b_1,b_2),\\
&& I_{\phi}(b_1,b_2)=\int\limits_0^{\pi}\frac{d\phi}{\pi} V \left(\sqrt{b_1^2+b_2^2-2b_1b_2\cos\phi}\right),\\
&& \rho_{00}=m_p^2\xi_1\xi_2,\; \rho_{0s}=m_p\xi_1,\; \rho_{s0}=m_p\xi_2,\; \rho_{ss}=1.
\end{eqnarray}
 
\begin{table}
\caption{\small\it\label{tab:3IP}Parameters of the model~\cite{3Pomerons}.}
\begin{center}
\begin{tabular}{|c|c|c|c|}
\hline
 $i$        &      1   & 2   & 3  \\
\hline
 $c_i$          &    $53.0\pm 0.8$   &   $9.68\pm 0.16$  &   $1.67\pm 0.07$ \\
\hline
  $r^2_i$ (GeV$^{-2}$)&     $6.3096\pm 0.2522$   &    $3.1097\pm 0.1817$ &   $2.4771\pm 0.0964$\\
\hline
\end{tabular}
\end{center}
\end{table}

\section*{Appendix B}

 For the calculation of elastic cross-sections we use the Bourrely-Soffer-Wu (BSW) parametrization~\cite{BSW}. Functions and values of parameters are given below. 
\begin{eqnarray}
&&\label{BSW0} {\rm T}(s,t_p)=\imath\int\limits_0^{\infty}b\; db\; J_0(b\sqrt{-t_p})(1-{\rm e}^{-\Omega_0(s,b)}),\\
&& \label{BSWel}\frac{d\sigma_{{\rm el}}}{dt_p}=\pi \left| {\rm T}(s,t_p)\right|^2,
\end{eqnarray}
\begin{eqnarray}
&& \label{BSWomega}\Omega_0(s,b)=\Omega_{IP}+\sum\limits_{i}\Omega_i,\\
&& \Omega_{IP}\simeq \frac{s^c}{\ln^{c^{\prime}}s}\left[1+\frac{{\rm e}^{\imath\pi c}}{\left( 1+\frac{\imath\pi}{\ln s}\right)^{c^{\prime}}} \right] F_{BSW}b\; {\rm for}\; s\gg m_p^2,|t|,
\end{eqnarray}

 For the $\pi^+ p$ elastic scattering we have $i=\rho$ in~(\ref{BSWomega}) and
\begin{eqnarray}
&& F^{\pi^+ p}_{BSW}(b)=\int\limits_0^{\infty}q\; dq\; J_0(q b) f_{\pi}\frac{a_{\pi}^2-q^2}{a_{\pi}^2+q^2}\frac{1}{(1+\frac{q^2}{m_1^2})(1+\frac{q^2}{m_2^2})(1+\frac{q^2}{m_{3\pi}^2})},\\
&& \Omega_{\rho}\simeq C_{\rho}(1+\imath)\left( \frac{s}{s_0}\right)^{\alpha_{\rho}(0)-1}\frac{{\rm e}^{-\frac{b^2}{4B_{\rho}}}}{2B_{\rho}},\\
&& B_{\rho}=b_{\rho}+\alpha_{\rho}^{\prime}(0)\ln\frac{s}{s_0},\; b_{\rho}=4.2704,\; \alpha_{\rho}(t)=0.3202+t,\; C_{\rho}=4.1624,
\end{eqnarray}
where values of parameters are listed in Table~\ref{tab:BSW}.

\begin{table}
\caption{\small\it\label{tab:BSW}Parameters of the model~\cite{BSW}.}
\begin{center}
\begin{tabular}{|c|c|c|c|c|c|c|c|c|}
\hline
 $c$   &  $c^{\prime}$   &  $m_1$   & $m_2$ & $m_{3\pi}$ & $f_{\pi}$ & $a_{\pi}$ & $f$ & $a$  \\
\hline
 0.167 &  0.748  & 0.577225  &  1.719896 & 0.7665 & 4.2414 & 2.3272 & 6.970913 & 1.858442 \\
\hline
\end{tabular}
\end{center}
\end{table}

For the $pp$ elastic scattering $i=\omega,\; A_2,\; \rho$ and
\begin{eqnarray}
&& F^{pp}_{BSW}(b)=\int\limits_0^{\infty}q\; dq\; J_0(q b)\; f\frac{a^2-q^2}{a^2+q^2}\frac{1}{(1+\frac{q^2}{m_1^2})^2(1+\frac{q^2}{m_2^2})^2},\\
&& \Omega_i(b)=\int\limits_0^{\infty}q\; dq\; J_0(q b)\; C_i {\rm e}^{-b_i q^2}(1\pm {\rm e}^{-\imath\pi\alpha_i(-q^2)}) \left( \frac{s}{s_0}\right)^{\alpha_i(-q^2)}.
\end{eqnarray}
Values of parameters are listed in Table~\ref{tab:BSWpp}.

\begin{table}
\caption{\small\it\label{tab:BSWpp}Parameters of the model~\cite{BSW} for secondary reggeons in $pp$ scattering.}
\begin{center}
\begin{tabular}{|c|c|c|c|}
\hline
 $i$   &  $\omega$   &  $A_2$   & $\rho$ \\
\hline
 $b_i,\; {\rm GeV}^{-2}$ & 0 & 0 & 8.54 \\
\hline
 $C_i$ & -167.3293 & -24.2686 & 124.91969 \\
\hline
 $\alpha_i(t)$ & $0.3229+0.7954t$ & $0.3566+t$ & $0.3202+t$ \\ 
\hline
\end{tabular}
\end{center}
\end{table}

 \noindent In this paper we take the following parametrization of the $\pi^+\pi^+$ elastic scattering, which is based on the BSW~\cite{BSW} one (approximate expressions for $s\gg s_0$, $|t|\ll 1$~GeV$^2$)
\begin{equation}
F^{\pi^+\pi^+}_{BSW}(b)\simeq\int\limits_0^{\infty}q\; dq\; J_0(q b)\; f_{\pi\pi}\frac{a_{\pi\pi}^2-q^2}{a_{\pi\pi}^2+q^2}\frac{1}{(1+\frac{q^2}{m_{3\pi}^2})^2},
\end{equation}
\begin{equation}
\Omega_0\simeq\Omega_{IP},\; f_{\pi\pi}=\frac{f_{\pi}^2}{f},\; \frac{1}{a_{\pi\pi}^2}=\frac{2}{a_{\pi}^2}-\frac{1}{a^2}.
\end{equation}

\section*{Appendix C}

Another parametrization for $pp$, $\pi^+ p$ and $\pi^+\pi^+$ cross-sections
is taken from~\cite{godizov1},\cite{godizov2}. The scattering amplitude is represented in the usual
eikonal form
\begin{equation}
\label{eikna}
T(s,b) = \frac{e^{2i\delta(s,b)}-1}{2i}
\end{equation}
(here $T(s,b)$ is the amplitude of the elastic scattering in the impact parameter 
$b$ space, $s$ is the invariant mass squared of colliding particles and
$\delta(s,b)$ is the eikonal function). Amplitudes in
the impact parameter space and momentum one are related thorough
the Fourier-Bessel transforms 
\begin{eqnarray}
&&\label{fourbess} f(s,b) = \frac{1}{16\pi s}\int_0^{\infty}d(-t)J_0(b\sqrt{-t})f(s,t)\,,\\
&& f(s,t) = 4\pi s\int_0^{\infty}db^2J_0(b\sqrt{-t})f(s,b)\,.
\end{eqnarray}

Eikonal function in the momentum space is
$$
\delta(s,t) = \delta_{\rm P}(s,t)+\delta_f(s,t)=
$$
\begin{equation}
\label{eikmin}
=\left(i+{\rm tg}\frac{\pi(\alpha_{\rm P}(t)-1)}{2}\right)
\beta_{\rm P}(t)\left(\frac{s}{s_0}\right)^{\alpha_{\rm P}(t)}+
\end{equation}
$$
+\left(i+{\rm tg}\frac{\pi(\alpha_f(t)-1)}{2}\right)
\beta_f(t)\left(\frac{s}{s_0}\right)^{\alpha_f(t)}
$$

The parametrization for the pomeron residue is 
\begin{equation}
\label{respom}
\beta_{\rm P}(t) = B_{\rm P}e^{b_{\rm P}\,t}
(1+d_1\,t+d_2\,t^2+d_3\,t^3+d_4\,t^4)\,,
\end{equation}
which is approximately (at low values of 
$d_1$, $d_2$, $d_3$ è $d_4$) an exponential at low $t$ values. Residues
of secondary reggeons we set as exponentials:
\begin{equation}
\label{ressec}
\beta_f(t) = B_fe^{b_f\,t}.
\end{equation}

\begin{table}[ht!]
\caption{\small\it\label{gtab1} Values of parameters of the model~\cite{godizov1},\cite{godizov2} for the $pp$ scattering.}
\begin{center}
\begin{tabular}{|l|l|l|l|l|l|}
\hline
& {\bf Pomeron}  &    & {\bf $f_2$-reggeon} &  & {\bf $\omega$-reggeon} \\
\hline
$p_1$ & $0.123$             & $c_f$ & $0.1$ GeV$^2$       & $c_\omega$ & $0.9$ GeV$^2$  \\
$p_2$ & $1.58$ GeV$^{-2}$    &       &                    &       &               \\
$p_3$ & $0.15$             &       &                    &       &              \\
$B_{\rm P}$ & $43.5$              & $B_f$  & $153$              & $B_\omega$ & $46$        \\
$b_{\rm P}$ & $2.4$ GeV$^{-2}$    & $b_f$  & $4.7$ GeV$^{-2}$   & $b_\omega$ & $5.6$ GeV$^{-2}$\\
$d_1$ & $0.43$ GeV$^{-2}$   &       &                    &       &               \\
$d_2$ & $0.39$ GeV$^{-4}$   &       &                    &       &               \\
$d_3$ & $0.051$ GeV$^{-6}$   &       &                    &       &              \\
$d_4$ & $0.035$ GeV$^{-8}$   &       &                    &       &              \\
\hline
$\alpha_{\rm P}(0)$ & $1.123$              & $\alpha_f(0)$ & $0.78$             
& $\alpha_\omega(0)$ & $0.64$ \\
$\alpha'_{\rm P}(0)$ & $0.28$ GeV$^{-2}$   & $\alpha'_f(0)$ & $0.63$ GeV$^{-2}$
& $\alpha'_\omega(0)$ & $0.07$ GeV$^{-2}$\\
\hline
\end{tabular}
\end{center}
\end{table}
\begin{table}[t!]
\caption{\small\it\label{gtab2} Values of parameters of the model~\cite{godizov1},\cite{godizov2} for the $\pi^{\pm} p$ scattering.}
\begin{center}
\begin{tabular}{|l|l|}
\hline
\multicolumn{2}{|c|}{$\pi^{\pm}\,p\to\pi^{\pm}\,p$}  \\
\hline
$B_{\rm P}$ & $26.7$  \\
$b_{\rm P}$ & $2.36$ GeV$^{-2}$ \\
$d_1$ & $0.38$ GeV$^{-2}$ \\
$d_2$ & $0.30$ GeV$^{-4}$ \\
$d_3$ & $-0.078$ GeV$^{-6}$ \\
$d_4$ & $0.04$ GeV$^{-8}$   \\
$B_f$ & $67$  \\ 
$b_f$ & $1.88$ GeV$^{-2}$ \\ 
\hline
\end{tabular}
\end{center}
\end{table}
Phenomenological parametrization for the "soft" pomeron trajectory is set to
\begin{equation}
\label{pomeron}
\alpha_{\rm P}(t) = 1+p_1\left[1-p_2\,t\left({\rm arctg}(p_3-p_2\,t)
                             -\frac{\pi}{2}\right)\right]\,.
\end{equation}
Trajectories of secondary reggeons $f_2$ and $\omega$ are parametrized by
functions
\begin{equation}
\label{second}
\alpha_{\rm R}(t) = \left(\frac{8}{3\pi}
\alpha_s(\sqrt{-t+c_{\rm R}})\right)^{1/2},\; {\rm R}=f,\omega,
\end{equation}
where
\begin{equation}
\label{analytic}
\alpha_s(\mu) = \frac{4\pi}{11-\frac{2}{3}n_f}
\left(\frac{1}{\ln\frac{\mu^2}{\Lambda^2}}
+\frac{1}{1-\frac{\mu^2}{\Lambda^2}}\right)
\end{equation}
is the one-loop analytic QCD running coupling~\cite{solovtsov}, $n_f = 3$ is the number of flavours, $\Lambda\equiv\Lambda^{(3)} = 0.346$~GeV~\cite{bethke}. Parameters 
$c_f\,,\;c_\omega>0$ are rather small to spoil the asymptotic behaviour
of secondary trajectories in the perturbative domain.

 Residues for $\pi\pi$, $\pi p$ and $pp$ scattering are assumed to be
\begin{equation}
\label{relproc}
\beta^{\pi\pi}_P(t)=\frac{\beta^{\pi p}_P(t)\beta^{\pi p}_P(t)}{\beta^{pp}_P(t)},
\end{equation}
\begin{equation}
\beta^{\pi\pi}_f(t)=\frac{\beta^{\pi p}_f(t)\beta^{\pi p}_f(t)}{\beta^{pp}_f(t)}.
\end{equation}

Parameters of the model are listed in Tables~\ref{gtab1},\ref{gtab2}.

\section*{Acknowledgements}

We are gratefull to M.~Albrow, V.I.~Kryshkin, N.E.~Tyurin, S.M.~Troshin for useful duscussions and helpfull suggestions, and also A.~Godizov for an alternative parametrization of
cross-sections and D.~Konstantinov for the help with modifications of the
software for Monte-Carlo simulation.

This work is supported by the grant RFBR-10-02-00372-a.

\end{document}